\documentclass[aps,prx,twocolumn,superscriptaddress,longbibliography]{revtex4-1}
	\usepackage[bookmarks=false,linkcolor=blue,urlcolor=blue,colorlinks,citecolor=blue]{hyperref}
	\usepackage{graphicx}
	\usepackage{color}
	\usepackage{amsmath}
	 \usepackage{setspace} 
	\usepackage[normalem]{ulem}
	\usepackage[toc,page]{appendix}

\begin{document}

\title{Quantum chaos in an electron-phonon bad metal}
	\date{\today}
	\author{Yochai Werman}
	\affiliation{Department of Condensed Matter Physics, Weizmann Institute of Science, Rehovot 76100, Israel}
\author{Steven A. Kivelson}
	\affiliation{Department of Physics, Stanford University, Stanford, California 93105, USA}
	\author{Erez Berg}
\affiliation{Department of Condensed Matter Physics, Weizmann Institute of Science, Rehovot 76100, Israel}
\affiliation{Department of Physics, James Franck Institute, University of Chicago, Chicago, IL 60637, USA}

\begin{abstract}
We calculate the scrambling rate $\lambda_L$ and the butterfly velocity $v_B$ associated with the growth of quantum chaos for a solvable large-$N$ electron-phonon system. 
We study a  temperature regime in which 
the electrical resistivity of this system exceeds the Mott-Ioffe-Regel limit and increases linearly with temperature -  
 a sign that there are no long-lived charged quasiparticles  - 
although the phonons remain well-defined quasiparticles. 
The long-lived phonons 
determine $\lambda_L$, rendering it parametrically smaller than the  theoretical upper-bound $\lambda_L \ll \lambda_{max}=2\pi T/\hbar$. 
Significantly, the chaos properties seem to be intrinsic - $\lambda_L$ and $v_B$ are the same for electronic and phononic operators. We consider two models - one in which the phonons are dispersive, and one in which they are dispersionless. 
In either case, we find that $\lambda_L$ is proportional to the inverse phonon lifetime, and $v_B$ is proportional to the effective phonon velocity. The 
thermal and chaos diffusion constants, $D_E$ and $D_L\equiv v_B^2/\lambda_L$, are 
 always comparable, $D_E \sim D_L$. In the dispersive phonon case, the charge diffusion constant $D_C$ satisfies $D_L\gg D_C$, while in the dispersionless case $D_L \ll D_C$.
\end{abstract}
\maketitle
	
\emph{Introduction.--} The thermalization properties of a closed interacting system have been characterized by the quantities governing many-body quantum chaos - the scrambling rate (or Lyapunov exponent) $\lambda_L$ and the butterfly velocity $v_B$\cite{Ovchinnikov, Sekino, Shenker, Kitaev, Patel, Banerjee, Chowdhury,Blake1}. These are defined by the intermediate-time increase in the expectation value of the squared commutator of two spatially separated local operators $A$ and $B$
\begin{equation}\label{eq:commutator}
\left\langle [A(0),B(x,t)][A(0),B(x,t)]^\dagger\right\rangle\sim e^{\lambda_L(t-\frac{|x|}{v_B})}.
\end{equation}
The butterfly velocity is the speed %in
at  which the effects of a local perturbation propagate, while the Lyapunov exponent $\lambda_L$ is a measure of the %time in 
rate at which the information of a local perturbation is scrambled into non-local degrees of freedom. It has been shown~\cite{MaldacenaBound} that in a general quantum system, $\lambda_L$ is bounded by the temperature, $T$ (setting $\hbar = k_B = 1$):
\begin{equation} \label{eq:bound}
\lambda_L\le \lambda_{max} = {2\pi T}.
\end{equation}

The notion of quantum chaos is especially pertinent to strongly interacting systems. The Lyapunov exponent saturates the bound in systems with a holographic dual~\cite{Shenker} and in the Sachdev-Ye-Kitaev model~\cite{SY, Kitaev, Maldacena, Stanford1}, and it is conjectured that %in 
 systems without long-lived ``quasiparticle'' excitations generically saturate the bound, i.e. $\lambda_L\sim T$~\cite{Patel}. Conversely, Fermi liquids~\cite{Aleiner, Banerjee} and weakly coupled large-$N$ models~\cite{Stanford1, Chowdhury} display a Lyapunov rate %which
 that  is parametrically smaller than this bound. 
%EBIt is therefore proposed that systems without quasiparticles can be distinguished from those with emergent long lived degrees of freedom by their universally large scrambling rate\cite{Patel}. 

$\lambda_L$ is generally not directly measurable in condensed matter systems. However, it was proposed that either the charge~\cite{Blake1, Blake2} or energy~\cite{Patel} diffusivities, $D_C$ and $D_E$ respectively, are related to $D_L \equiv v_B^2/\lambda_L$. 
%, , it has been proposed~\cite{Blake1, Blake2, Patel} that quantum chaos governs the thermal transport properties of strongly correlated systems through the relation
%\begin{equation}
%D_E = \frac{v_B^2}{\lambda_L},
%\end{equation}
%where $D_E$ is the thermal diffusion coefficient.

A well-known class of strongly correlated condensed matter systems are the ``bad metals"~\cite{Emery, Hussey}. These materials, which vary microscopically, are defined by a metallic resistivity which increases linearly with temperature, and reaches values above the Mott-Ioffe-Regel (MIR) limit~\cite{Ioffe} without saturation. The large value of resistivity in these materials at high temperatures suggests non-quasiparticle transport and rapid momentum degradation, the characteristics of an incoherent metal. 
%EBIt has been suggested that these materials are governed by a universal dissipative timescale $\tau\sim 1/T$\cite{SubirBook, Bruin}. Eq.~(\ref{})
%, and that they saturate a conjectured universal bound on the charge diffusion constant
%\begin{equation}
%D_C \ge \frac{v^2}{T},
%\end{equation} 
%with $v$ a microscopic velocity scale~\cite{Hartnoll}. It has lately been argued that this scale is the butterfly velocity, $v_B$~\cite{Blake1, Blake2}. 

The strongly coupled nature and anomalous transport properties of bad metals naturally raise the question of chaos in these systems. In particular, does the bound on the scrambling rate, Eq.~(\ref{eq:bound}), lead to a universal bound on transport in bad metals~\cite{Hartnoll}? Is a strongly incoherent metal necessarily also strongly chaotic, in the sense that Eq.~(\ref{eq:bound}) is saturated? In this paper, we calculate the Lyapunov exponent, butterfly velocity, and thermal diffusion coefficients for a class of solvable large-$N$, electron-phonon systems which display bad metallic behavior - namely, a linear-in-$T$ growth of the electrical resistivity above the MIR limit. We find that the scrambling rate and the butterfly velocity are the same whether one uses electron or phonon operators to define them, suggesting that $\lambda_L$ and $v_B$ are intrinsic properties of the system. %EBindependent of the operators used. 
The results are formulated in Table \ref{results}. Our main result is that scrambling and energy are both transported mainly by the phonons, in contrast with charge transport. We therefore find that $D_E$ is parametrically the same as $D_L$, 
$\lambda_L$ is far from saturating the bound in Eq.~(\ref{eq:bound}), %while 
and $D_L$ bears no relation to the charge diffusion constant, $D_C$.

\emph{Model.--}
Our system is composed of $N\gg1$ electron bands, which interact with $N^2$ optical phonon modes, on a $d$-dimensional lattice. On each site, the phonon displacement couples to the electron density. In this type of large-$N$ expansion, inspired by the work of Fitzpatrick et al.~\cite{Raghu}, the large number of phonons strongly renormalize the electrons, as reflected by an $O(1)$ electron self energy. On the other hand, the backaction of the electrons on the phonons
and on the electron-phonon vertex is suppressed by a factor of $1/N$.

%\begin{widetext}
\begin{center}
\begin{table*}[]
\centering
\caption{Comparison of the diffusion coefficients of charge,
%Chaos quantities, charge diffusion 
 $D_C$, % and 
 heat,
 % diffusion (
 $D_E$ %
 and chaos, $D_L\equiv v_B^2/\lambda_L$ %coefficients
  for the two large-$N$ electron-phonon models studied - that with dispersive and non-dispersive phonons.  Here $v_B$ is the ``butterfly velocity'' and $\lambda_L$ is the scrambling rate which in all cases is parametrically smaller than the bound $2\pi T$,   $\omega_0$ is the %phonon Einstein frequency in the non-dispersive model, and a representative frequency in the dispersive model. 
  average phonon frequency, $E_F$ is the electron Fermi energy, and $g$ %$\lambda$ 
  is the dimensionless electron-phonon coupling. $v_{el}$ and $v_{ph}$ are, respectively, the  root mean squared bare  electron and phonon velocities (averaged over the Brillouin zone), where $v_{ph} = 0$ in the dispersionless model. %The scrambling rate $\lambda_L$ is parametrically smaller than the bound $2\pi T$.
  All results assume $N$ large and $\omega_0 \ll T \ll E_F \ll g T$.
  }
\label{results}
\begin{tabular}{|c|c|c|c|c|c|}
\hline
                                       &  $\lambda_L$                                  & $v_B$                                                                 & $D_C$                  & $D_E \sim D_L$\\  \hline
Dispersionless phonons & $\frac{1}{N}\frac{\omega_0^2}{T}$ & $\frac{1}{N}\frac{\omega_0}{T}\sqrt{\frac{E_F}{g T}}v_{el}$  & $ \frac{v_{el}^2}{\sqrt{E_Fg T}}$ & $\frac{1}{N}\frac{E_F}{g T}\frac{v_{el}^2}{T}$ \\  \hline
Dispersive phonons       & $\frac{1}{N}\frac{\omega_0^2}{T}$ & $v_{ph}$    & $ \frac{v_{el}^2}{\sqrt{E_Fg T}}$ & $N{v_{ph}^2}\frac{T}{\omega_0^2}$ \\ \hline
\end{tabular}
\end{table*}
\end{center}
%\end{widetext}

The Hamiltonian is given by
\begin{equation}\label{eq:model}
H = H_{\mathrm{el}} + H_{\mathrm{ph}} + H_{\mathrm{int}},
\end{equation}
where
\begin{eqnarray}
H_{\mathrm{el}}&=&\sum_{a=1}^N\frac{ d^dk}{(2\pi)^d}\xi_{\mathbf{k}} c^\dagger_a(\mathbf{k})c_a(\mathbf{k}), \nonumber \\
H_{\mathrm{ph}} &=& \sum_{a,b=1}^N \sum_{\omega_n}\int \frac{d^dq}{(2\pi)^d}\left[\frac{1}{2}M\omega_\mathbf{q}^2|X(\mathbf{q})|^2+\frac{|P(\mathbf{q})|^2}{2M}\right], \nonumber \\
H_{\mathrm{int}} &=& \frac{\alpha}{\sqrt{\beta N}}\sum_{a,b=1}^N\sum_iX_{ab}(i)\left[c^\dagger_a(i)c_b(i)+a\leftrightarrow b\right].
\end{eqnarray}

%\begin{equation}
%H_{\mathrm{el}}=\sum_{a=1}^N\frac{ d^dk}{(2\pi)^d}\xi_{\mathbf{k}} c^\dagger_a(\mathbf{k})c_a(\mathbf{k})
%\end{equation}
%is the electronic part,
%\begin{equation}
%H_{\mathrm{ph}} = \sum_{a,b=1}^N \sum_{\omega_n}\int \frac{d^dq}{(2\pi)^d}\left[\frac{1}{2}M\omega_\mathbf{q}^2|X(\mathbf{q})|^2+\frac{P(\mathbf{q})^2}{2M}\right]\end{equation}
%is the phononic part, and
%\begin{eqnarray} \label{eq:Lint}
%H_{\mathrm{int}} &=& \frac{\alpha}{\sqrt{\beta N}}\sum_{a,b=1}^N\sum_iX_{ab}(i)\left[c^\dagger_a(i)c_b(i)+a\leftrightarrow b\right]
%\end{eqnarray}
%is the electron-phonon interaction term.
%~\footnote{In the electron-phonon coupling, we keep only linear order in the phonon displacement, and neglect higher-order terms. This is justified, even at high temperature, in the large-$N$ limit; this is since the typical magnitude of the coupling term to any single phonon mode is $\alpha \sqrt{T/(K\, N)}$, which is smaller than other electronic scales (e.g., $E_F$).}.%\cite{footnote-linear}.
Here, $c^\dagger_a(\mathbf{k})$ is the Fourier transform of $c^\dagger_a(i)$, which creates an electron of flavor $1 \le a \le N$ on site $i$; the electronic dispersion is $\xi_{\mathbf{k}}=\epsilon(\mathbf{k})-\mu(T)$, with $\epsilon(\mathbf{k})\in [-\Lambda/2,\Lambda/2]$ (where $\Lambda$ is the bandwidth). $\mu(T)$ is the chemical potential 
%at temperature $T$, 
and $\beta=1/T$. 
%EB
For simplicity, we take $\epsilon{(\mathbf{k}})$ to correspond to a nearest-neighbor hopping on a $d-$dimensional hypercubic lattice. Furthermore, we will work at half filling, so that $\mu(T) = 0$ and the Fermi energy, $E_F=\Lambda/2$. None of the results in the relevant range of $T$ depend qualitatively on these choices.
$X_{ab}(\mathbf{q})$ is the Fourier transform of the phonon displacement operator $X_{ab}(i)$ of flavor $a,b$, and $P_{ab}$ is the conjugate momentum; $M$ is the ionic mass, and $\alpha$ is the electron-phonon coupling strength. The lattice spacing $a$ is set to $1$. 

We consider two types of models which differ by the $\mathbf{q}$ dependence of the phonon frequency $\omega_\mathbf{q}$; a dispersionless (Einstein) model and a model with dispersive optical phonons.  In all cases, we define the typical phonon frequency to be 
%with 
$\omega_0\equiv  \left\langle \omega_\mathbf{q}\right\rangle$, %$\omega_\mathbf{q}>0$, 
and the root-mean-squared phonon velocity as, $v_{ph} =\sqrt{\left\langle \left|\partial_{\mathbf{q}}\omega_\mathbf{q}\right|^2\right\rangle}$ where the brackets denote an average over the Brillouin zone.  %\omega_\mathbf{q} = \omega_0$, and a model with dispersive optical phonons.  such that $\omega_0 = \left\langle \omega_\mathbf{q}\right\rangle$, $\omega_\mathbf{q}>0$, with the brackets denoting an average over the Brillouin zone. 
($v_{ph}=0$ in the non-dispersive model.)

We define the dimensionless electron-phonon coupling constant (which we will take to be large)
\begin{eqnarray}
\label{eq:c}
%\lambda=
g\equiv \frac{\alpha^2\nu}{M\omega_0^2}\gg 1,
\end{eqnarray}
with $\nu$ the density of states at the Fermi energy. As in Ref.~\cite{Werman, Werman2}, we consider temperatures such that $\omega_0 \ll T \ll E_F \ll gT$, %where  $E_F$ is the Fermi energy (measured from the band bottom), 
%and focus on the regime $%\lambda 
%g T\gg E_F$
 - a ``high temperature''  regime in which  electronic quasiparticles are no longer well-defined. %EBscattering rate is larger than the electronic energy scale.

\emph{Single particle properties.--} Taking the limit $N\rightarrow \infty$ allows us to solve the model (\ref{eq:model}) order by order in $1/N$. In the rest of this section, we present results for the dispersionless phonons, while the dispersionful case is considered in the appendix. 
Just as in \cite{Werman}, the leading order contributions result in a self-consistent Dyson's equation for the fermion self-energy $\Sigma(\omega)$ (more details appear in the appendix):
\begin{eqnarray}\label{eq:selfenergy}
\Sigma(\omega) &=&\frac{%\lambda
g T}{\nu}\int\frac{d^dq}{(2\pi)^d}\frac{1}{\omega - \xi_{\mathbf{q}}-\Sigma(\omega)}.
\label{eq:sigma}
\end{eqnarray}

The electron self-energy in the high-$T$ regime $%\lambda
g T\gg E_F$ is given by
\begin{equation}
\Sigma(\omega) =\frac{1}{2} \left[\omega-i\sqrt{\frac{4%\lambda 
gT}{\nu}-\omega^2}\right];
\end{equation}
Note that in this regime $\Sigma''(0)\gg E_F$, i.e., there are no well-defined electron-like quasiparticles. 
%EBthe large scattering rate on the Fermi surface, $\Sigma''(0)\gg E_F$, and the fact that the spectral function $A_\mathbf{k}(\omega) = -2\Im[G_\mathbf{k}(\omega)]  \approx \frac{\nu}{\lambda T}\sqrt{\frac{4\lambda T}{\nu}-\omega^2}$, with $G_\mathbf{k}(\omega)\sim[\omega-\Sigma(\omega)]^{-1}$ the electronic Green's function, is spread over a frequency interval larger than the Fermi energy, render the electronic quasiparticle ill-defined. 

The phonon self energy is subleading in $1/N$; however, it is of crucial importance in calculating the scrambling rate. To leading order in $1/N$, it is given by the fermion polarization function: %EB, which corresponds to (more details appear in the appendix)
\begin{eqnarray}
\Gamma(\mathbf{q},i\omega_n) &=& -\frac{\alpha^2}{NM\omega_0}\frac{1}{\beta}\sum_{\nu_n}\\
&\times&\int \frac{d^dk'}{(2\pi)^d}\mathcal{G}(\mathbf{k'},i\nu_n)\mathcal{G}(\mathbf{k'+q},i\nu_n+i\omega_n)\nonumber
\label{Gamma}
\end{eqnarray}

The phonon lifetime is given by $\Gamma''(\mathbf{q},q_0) = -\Im\Gamma^R(\mathbf{q},q_0)$, 
which for small real external frequencies $q_0\ll T$ results in (see Appendix)
\footnote {It is at first surprising that the phonon decay rate in Eq.~(\ref{phononlifetime}) is independent of $g$, given that it must vanish as $g\to 0$;  however, here the explicit proportionality to $g$ from Eq.~(\ref{Gamma}) is cancelled by an implicit dependence of the electron-response function in the assumed limit $g\gg E_F/T$.}

\begin{eqnarray}
\Gamma''(\mathbf{q},q_0) &=& -\frac{1}{4\pi N}\frac{q_0\omega_0}{T}
\label{phononlifetime}
\end{eqnarray}

As we are concerned with the spatial propagation of scrambling, we %will 
 also need the velocity of the phonons. %In the case where 
 Because the bare phonon dispersion vanishes, the effective phonon dispersion originates in the $\mathbf{q}$ dependence of the real part of the self-energy, $\Gamma'(\mathbf{q},\omega) = \Re\Gamma^R(\mathbf{q},\omega)$; for $\omega=0$, this becomes to lowest order in $E_F/%\lambda
g T$

\begin{equation}
\tilde{v}_{ph}(\mathbf{q})%=
\equiv\partial_\mathbf{q}\Gamma'(\mathbf{q},%\omega=
0) =\frac{1}{2\pi N}\frac{\omega_0}{T}\sqrt{\frac{\nu}{%\lambda
g T}}\int \frac{d^dk}{(2\pi)^d} \epsilon_\mathbf{k}v_\mathbf{k+q}.
\label{phonondispersion}
\end{equation}

\emph{Electronic scrambling rate.-- } Following Ref.~\cite{Stanford}, we calculate a regularized version of Eq.~(\ref{eq:commutator}), in which the thermal density matrix ${\rho} = e^{-\beta H}/\mathrm{Tr}[e^{-\beta H}]$ is split between the two commutators. We define the quantity
%EBWe first calculate the scrambling rate of the electrons; we define the the flavor- and position- averaged quantity
\begin{eqnarray}\label{eq:f}
&& \tilde f(x,t)  = \frac{\theta(t)}{N^2} \\
&& \times \sum_{ab} \mathrm{Tr}\left[\sqrt{\rho}\{c_a(x,t),c^\dagger_b(0,0)\}\sqrt{\rho}\{c_a(x,t),c^\dagger_b(0,0)\}^\dagger\right]. \nonumber
\end{eqnarray}
 $f(\mathbf{q},\omega) = \int d^dx dt e^{-i(\mathbf{q}\mathbf{x}-\omega t)} \tilde f(x,t)$ is the Fourier transform of (\ref{eq:f}) in space and its Laplace transform in time. 
%EBAt intermediate times, we expect that $f(x, t)$ to follow the behavior . 
To proceed, we evaluate Eq.~(\ref{eq:f}) on a generalized Keldysh contour (see Ref.~\cite{Stanford}, and the Appendix). 
%EBThe Laplace transform of this function, $f(\omega)$, 
$f(\mathbf{q}, \omega)$ obeys the Bethe-Salpeter equation (schematically represented in Fig.~\ref{fig:elchaos})

\begin{eqnarray}\label{eq:Bethe2}
&&f(\mathbf{q},\omega) = \frac{1}{N}\int \frac{d^{d+1}}{(2\pi)^{d+1}}k\,f(k,\mathbf{q},\omega)\\
&&f(k,\mathbf{q},\omega) = G^R(k+\mathbf{q})G^A(k-\omega)\nonumber\\
&&\times\left[1-\alpha^2\int \frac{d^{d+1}k'}{(2\pi)^{d+1}} D^W({k-k'})f(k',\mathbf{q},\omega)\right.\nonumber\\
&&+\left.\frac{1}{N}\alpha^4\int \frac{d^{d+1}k'}{(2\pi)^{d+1}} \mathcal{F}(k,k',\mathbf{q},\omega)f(k',\mathbf{q},\omega)
\right]+O\left(\frac{1}{N}\right).\nonumber
\end{eqnarray}
%EBHere
%\begin{eqnarray}\label{Eq.F}
%\mathcal{F}(k,k',\mathbf{q},\omega) &=& \int d^{d+1}k_1 D^R(k_1+\mathbf{q})D^A(k_1-\omega)\nonumber\\
%&\times& G^W({k-k_1})G^W({k'-k_1}),\nonumber\\
%\end{eqnarray}

\begin{figure}
  \centering
    \includegraphics[width=0.4\textwidth]{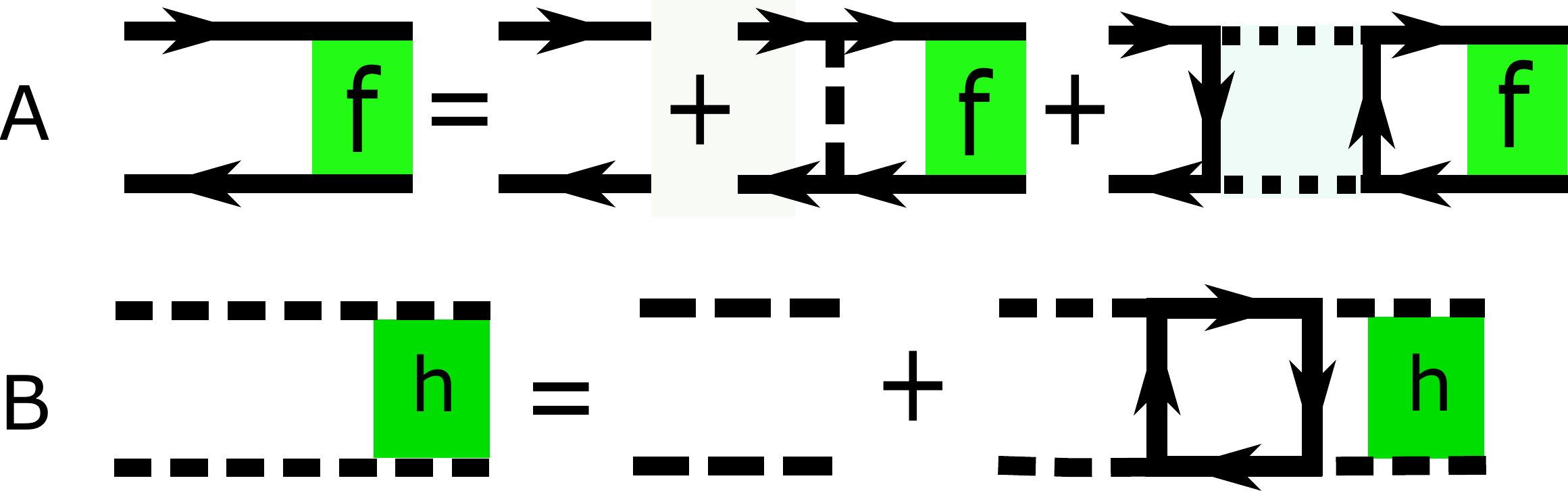}
  \caption{The Bethe-Salpeter equation for A) the electronic $f(\omega)$ B) the phononic $g(\omega)$. Horizontal lines represent retarded/advanced propagators, with the top line living at the upper branch ($i\beta/2$) and the bottom line on the lower branch. Vertical lines are Wightman propagators. The last diagram on each row is naively suppressed by $1/N$, but diverges in the limit $\tau_{ph}\rightarrow\infty$.}
\label{fig:elchaos}
\end{figure}

Here, $k$ is a $d+1$-vector $(k_0, \mathbf{k})$, and
\begin{eqnarray}\label{Eq.F}
\mathcal{F}(k,k',\omega) &=& \int \frac{d^{d+1}k_1}{(2\pi)^{d+1}} D^R(k_1)D^A(k_1-\omega)\\
&&\times G^W({k-k_1})G^W(k'-k_1)\nonumber
\end{eqnarray}
and we define
\begin{eqnarray}
&&\frac{1}{Z}\mathrm{Tr}\left[e^{-\beta H} X_{\alpha\beta}(x,t)X_{\alpha\beta}(0,i\frac{\beta}{2}) \right]\equiv D^W(x,t)
\end{eqnarray}
as the phonon Wightman propagator, and in the same way 
\begin{eqnarray}
&&\frac{1}{Z}\mathrm{Tr}\left[e^{-\beta H} c_{a}(x,t)c^\dagger_{a}(0,i\frac{\beta}{2}) \right] \equiv G^W(x,t).
\end{eqnarray}
The $\alpha^4$ term in Eq.~(\ref{eq:Bethe2}) is naively suppressed by a factor of $1/N$ relative to the rest.
 However, %if it is 
 when evaluated with the bare phonon propagators, it diverges; one must include the phonon self-energy, which renders this term of the same order as the other diagrams.

If $\tilde f%(t)
$ has an exponentially growing part, $\tilde f(\mathbf{0}, t) = \frac{1}{N}e^{\lambda_L t}+O\left(\frac{1}{N^2}\right)$, its Laplace transform $f(\mathbf{q},\omega)$ should divrege for $\omega = i\lambda_L$. We look for divergencies in $f(\mathbf{k},k_0,\omega=i\lambda_L)$, and for this imaginary frequency we may ignore the inhomogeneous term in Eq.~(\ref{eq:Bethe2}). Solving the resulting equation for $\omega = i\lambda_L$ results in
\begin{equation}
\lambda_L \sim 0.25 \frac{1}{N}\frac{\omega_0^2}{T}.
\end{equation}

%EBWe expand this equation in small $q/k_F$; 
The butterfly velocity $v_B$ can be extracted by the relation $[f(\mathbf{q},\lambda_L)-f(q=0,\lambda_L)]/f(q=0,\lambda_L) \sim v_B^2q^2/\lambda_L^2$ (more information appears in the appendix). This results in
\begin{equation}
v_B\sim\max\left(v_{ph}, \partial_\mathbf{q}\Gamma'(\mathbf{q},\omega=0)\right),
\end{equation}
i.e. the butterfly velocity is equal to the characteristic group velocity of the phonons.

%EB\emph{Butterfly velocity.-- }The out-of-time-ordered correlation function, evaluated at two spatially separated points, allows us to characterize the evolution of scrambling in both space and time. We therefore consider the function $f(x,t)$, defined in Eq.~(\ref{eq:f}), and calculate its Fourier transform in space and Laplace transform in time, $f(\mathbf{q},\omega) = \int d^dx dt e^{-i(\mathbf{q}\mathbf{x}-\omega t)} f(x,t)$; this function obeys the $\mathbf{q}$-dependent Bethe-Salpeter equation

%\begin{eqnarray}\label{eq:Bethe2}
%&&f(\mathbf{q},\omega) = \frac{1}{N}\int d^{d+1}k\,f(k,\mathbf{q},\omega)\\
%&&f(k,\mathbf{q},\omega) = G^R(k+\mathbf{q})G^A(k-\omega)\nonumber\\
%&&\times\left[1-\alpha^2\int d^{d+1}k' D^W({k-k'})f(k',\mathbf{q},\omega)\right.\nonumber\\
%&&+\left.\frac{1}{N}\alpha^4\int d^{d+1}k' \mathcal{F}(k,k',\mathbf{q},\omega)f(k,\mathbf{q},\omega)
%\right]+O\left(\frac{1}{N}\right).\nonumber
%\end{eqnarray}
%
%Here
%\begin{eqnarray}\label{Eq.F}
%\mathcal{F}(k,k',\mathbf{q},\omega) &=& \int d^{d+1}k_1 D^R(k_1+\mathbf{q})D^A(k_1-\omega)\nonumber\\
%&\times& G^W({k-k_1})G^W({k'-k_1}),\nonumber\\
%\end{eqnarray}

\emph{Phonon scrambling rate.--} A similar calculation produces the scrambling rate and butterfly velocity for the phonons, obtained from the quantity 
\begin{eqnarray}
&&h(x,t) = \frac{1}{N^4}\theta(t)\sum_{abcd}\\
&&\mathrm{Tr}\left[\sqrt{\rho}\left[X_{ab}(x,t),X_{cd}(0,0)\right]\sqrt{\rho}\left[X_{ab}(x,t),X_{cd}(0,0)\right]^\dagger\right].\nonumber
\end{eqnarray}
It results in the same values for $\lambda_L$ and $v_B$, which suggests that the chaos quantities are intrinsic to the model, independent of which operators are chosen.

\emph{Thermal transport- }The Matsubara thermal current operator of this system is given by (see appendix for details)
\begin{equation}
J^Q(i\omega_n) =\frac{1}{\beta}\sum_{a,\nu_n}\nu_n\int \frac{d^dk}{(2\pi)^d} v_{el}(\mathbf{k})c_a^\dagger(\mathbf{k},\nu_n)c_a(\mathbf{k},\nu_n+\omega_n);
\end{equation}
we calculate the thermal conductivity following the Luttinger prescription~\cite{Luttinger1, Luttinger2, Shastry}
\begin{equation}
\kappa = -\frac{1}{T}\lim_{\omega\rightarrow 0} \frac{\Im\left[\Pi(\omega)\right]}{\omega},
\end{equation}
where $\Pi(\omega)$ is the retarded thermal current-thermal current correlation function. To leading order in $1/N$ and $E_F/g T$, this results in
\begin{equation}
\kappa =  \left \{
  \begin{tabular}{cc}
  $N^3\langle v_{ph}^2\rangle \frac{T}{\omega_0^2}$ & (Dispersive case)   \\
  $N^3\langle \tilde{v}_{ph}^2\rangle \frac{T}{\omega_0^2}$ & (Non-dispersive case)   
  \end{tabular}
\right.
\end{equation}

In both cases, the heat capacity is dominated by the phonons, $c=N^2k_B$ (where we have inserted $k_B$ for clarity). Using the Einstein relation $\kappa = cD_E$, we arrive at the thermal diffusion coefficients.

\emph{Charge transport.--} The conductivity of this model has been analyzed in Refs.~\cite{Werman, Werman2}. In the high temperature regime $g T\gg E_F$, the resistivity is given by
\begin{equation}
N\rho(T) = \frac{g T}{\nu\langle v_{el}^2\rangle},
\end{equation}
which parametrically exceeds the Mott-Ioffe-Regel limit~\cite{Ioffe}, which is $N\rho_{\mathrm{MIR}} = O(1)$ in our units~\footnote{In conventional units, $N\rho_{MIR} = \hbar a^{d-2}/e^2$}. %EB, reflecting the fact that the electronic scattering rate is much larger than the bandwidth.

It should be emphasized that there momentum is not even nearly conserved in our model; in the regime $T\gg\omega_0$, phonons throughout the Brillouin zone are excited, so that umklapp scattering is just as efficient as normal scattering. In addition, we show in the appendix that phonon drag contributions are suppressed in this model, leading to a large resistivity.

\emph{Discussion.--} We calculate the scrambling rate $\lambda_L$ and butterfly velocity $v_B$ for both the electrons and the phonons in this system. We find that they are the same for the two operators - which suggests that $\lambda_L$ and $v_B$ are intrinsic properties, independent of the specific operators used in Eq.~(\ref{eq:commutator}). Intuitively, since the electrons and phonons are coupled, it is expected that scrambling (i.e., the spread of operators over the system) of one would lead to scrambling of the other; therefore, we expect that there cannot be two independent scrambling rates. 
This is consistent with the finding of Ref.~\cite{Chowdhury} that the scrambling rate in an $O(N)$ field theory is the same of the primary field and for a composite operator. 

Our system is composed of strongly renormalized electrons coupled to weakly interacting phonons. The phonon excitations are well defined quasiparticles - in the sense that lifetime $\tau = NT/\omega_0^2$ is parametrically longer than the inverse energy, $1/\omega_0$. This explains why the bottleneck for scrambling in this system is the loss of phonon phase coherence, and thus the scrambling rate is determined by the phonon scattering rate. This is also consistent with the fact that in the limit $\omega_0\rightarrow 0$, where the phonons act essentially as (annealed) disorder, the scrambling rate must vanish~\cite{Swingle2017,Patel2}. Similarly, we find that the butterfly velocity is proportional to the effective {phonon} velocity, and is unrelated to the electronic velocity.

The scrambling rate we find is far from the bound, $\lambda_L\ll \lambda_{max}$, due to the existence of phononic quasiparticles. Nevertheless, charge transport in the system does not reflect this fact - the resistivity of this model is larger than the Ioffe-Regel limit, due to the strong scattering of the electrons. This supports the notion that in general, $D_C$ is unrelated to $D_L$~\cite{Davison}. 
That $D_L\approx D_E$ is  consistent with the fact that in our model both the thermal transport and the propagation of chaos are dominated by  phonons. 
%EB, as the scram is carried mainly by the phonons. In contrast, thermal transport is dominated by the phonons, even if the phonons have no bare dispersion. 
%EBTherefore, in our system, $D_L\approx D_E$.

These results raise the question of the general relation between energy diffusion and scrambling. Clearly, the energy diffusivity can be much larger than $D_L$; this is the case, for instance, in a relativistic field theory where momentum is conserved~\cite{Chowdhury}. It is therefore plausible to propose that
%As is well known, in systems where the momentum relaxation rate $\Gamma_P$ is the smallest rate in the system, $D_E\propto 1/\Gamma_P$ will be much larger than $D_L$. It is therefore plausible to propose that for translationally invariant systems,
\begin{equation} \label{eq:bound_DE}
D_E\ge D_L.
\end{equation}
This relation does not hold universally; a one-dimensional counter example has been presented in Ref.~\cite{Gu}. The extent to which Eq.~(\ref{eq:bound_DE}) is valid in spatial dimensions $d>1$ 
%and in (statistically) homogeneous systems, 
remains an open question. 
Moreover, even concerning the relation between $D_C$ and $D_L$, it remains an open question whether in a system in which all degrees of freedom are strongly scattered - one in which $\lambda_L \sim 2\pi T$ - a more general relation between $D_C$ and $D_L$ might be possible.  For instance, in Ref. \cite{Kapitulnik}, evidence was  presented that in the normal state of the cuprates, $D_C \sim D_E$, and furthermore it was speculated that both are comparable to the bounding value of $D_L$ with $\lambda_L \sim 2\pi T$.
%EBThe fact that charge transport in this system is decopled from the quantum chaos properties, and furthermore that the electronic scattering rate may be arbitrary large, due to the coupling to the large phononic bath, seems to imply that charge transport, which is only carried by the electrons, can not be bound. On the other hand, thermal transport, which is carried by all excitations in the system, may be bound.

%\bibliographystyle{plainnat}

\emph{Acknowledgements.--} We thank  Ehud Altman, Debanjan Chowdhury, Sean Hartnoll, Andy Lucas, Aavishkar Patel, and Subir Sachdev for enlightening discussions. SAK was supported by the Department of Energy, Office of Basic Energy Sciences under contract DE-AC02-76SF00515.

\begin{widetext}

\section*{Supplementary Material for: \\ Quantum chaos in an electron-phonon bad metal}

\section{Model}
\subsection{Hamiltonian}

The system we consider is composed of $N\gg1$ identical electron flavors interacting with $N^2$ optical, phonon modes, in $d$ spatial dimensions. In this type of large-$N$ expansion, inspired by the work of Fitzpatrick et al.~\cite{Raghu}, the phonon modes act as a momentum and energy bath for the electrons, while each phonon mode is only weakly affected by the electrons. 

The Hamiltonian of the system is given by
\begin{equation}\label{eq:model}
H = H_0+ H_{\mathrm{int}} \equiv H_{\mathrm{el}} + H_{\mathrm{ph}} + H_{\mathrm{int}},
\end{equation}
where
\begin{equation}
\begin{split}
H_{\mathrm{el}}&=\sum_{a=1}^N\int\frac{ d^dk}{(2\pi)^d} \xi_{\mathbf{k}} c^\dagger_a(\mathbf{k}) c_a(\mathbf{k})
\\
H_{\mathrm{ph}} &= \sum_{a,b=1}^N \int \frac{d^dq}{(2\pi)^d}\left[\frac{1}{2}M\omega_{\mathbf{q}}^2|X_{ab}(\mathbf{q})|^2 + \frac{1}{2M}|P_{ab}(\mathbf{q})|^2\right],\\
H_{\mathrm{int}} &= \frac{\alpha}{\sqrt{N}}\sum_{a,b=1}^N \int\frac{d^dk}{(2\pi)^d}\frac{ d^dq}{(2\pi)^d}X_{ab}(\mathbf{q})c^\dagger_a(\mathbf{k}) c_b(\mathbf{k+q}).
\end{split}
\end{equation}

The electronic Hamiltonian is given in terms of operators $c^\dagger_a(\mathbf{k})$ which create an electron of flavor $a$ and energy $\xi_{\mathbf{k}}$. For simplicity, we consider an particle-hole symmetric spectrum and work at an electronic density of $n=0.5$ per flavor, which allows us to set the chemical potential $\mu(T) $ to $0$ at all temperatures. 

$X_{ab}$ is a symmetric matrix of phonon displacement operators, and $P_{ab}$ the corresponding momenta. We consider an optical spectrum $\omega_\mathbf{q} = \omega_0+\omega_1(\mathbf{q})$. To simplify the calculations, we assume that $\omega_0\gg |\omega_1(\mathbf{q})|$; 
%{\color{red} Is this necessary? $\omega_0 > |\omega_1(q)|$ is not sufficient?} 
thus the only effect of $\omega_1(\mathbf{q})$ is the introduction of a finite bare phonon velocity
\begin{equation}
v_{ph}^2 = \left\langle \left(\partial_\mathbf{q}\omega_\mathbf{q}\right)^2\right\rangle_{BZ},
\end{equation}
where the brackets denote an average over the first Brillouin zone (BZ). We consider cases in which $v_{ph}$ is either zero or non-zero, as explained below. For concreteness, we will use $\omega_1(\mathbf{q}) = \omega_1 \sum_{i=1\dots d}\cos(q_i)$ when it is needed in the calculation.

We define the dimensionless coupling constant
\begin{eqnarray}
g = \frac{\alpha^2\nu}{M\omega_0^2}\gg 1,
\end{eqnarray}
with $\nu$ the $T=0$ electronic density of states at the Fermi energy.

%EBOur system is composed of two distinct species - electrons and phonons. 
The coupling between each flavor of phonon to the electrons is suppressed by $1/\sqrt{N}$; the relatively small number of electrons relative to the phonons thus implies that the phonons are weakly affected by the interactions, resulting in self-energy corrections that appear only at subleading orders in $1/N$. The electrons, on the other hand, are strongly affected by the $N^2$ phonos, and their self energy can become larger than the Fermi energy, $E_F$. This is a consequence of our particular large-$N$ limit.

We consider temperatures such that
\begin{eqnarray}
\omega_0\ll T\ll E_F \lesssim  W,
\end{eqnarray}
with $W$ the bandwidth. We focus on the high temperature regime $gT\gg W$, in which the electronic lifetime is shorter than $W^{-1}$.

\subsection{Self energies}\label{selfenergies}

\subsubsection{Electron self energy}
Just as in \cite{Raghu}, the full set of rainbow diagrams, depicted in Figure~\ref{fig:rainbow}, contributes to the electron self-energy to lowest order in $1/N$. This results
in a self-consistent Dyson's equation for the fermion self-energy:
\begin{eqnarray}\label{eq:selfenergy}
\Sigma(\mathbf{k},{k_0}) = \Sigma({k_0})=-\frac{\alpha^2 T}{M}\int\frac{d^dq}{(2\pi)^d}\frac{1}{\omega_\mathbf{q}^2}\frac{1}{{-k_0} + \xi_{\mathbf{k+q}}+\Sigma(k_0)}.
\label{eq:sigma}
\end{eqnarray}
where $k_0$ denotes the frequency. Here, for simplicity, we consider a model with a momentum independent electron-phonon coupling; 
%EBthe limit where the phonon dispersion is almost flat; 
as a result, the electron self-energy at high temperatures is nearly momentum independent, as we will show. We do not expect this assumption to qualitatively change the final results for the scrambling rate and the thermal conductivity; as shown previously, the charge transport at high temperatures is sensitive to the nature of the electron-phonon coupling~\cite{Werman2}. 

At high temperatures $gT \gg W$, we neglect the electronic dispersion relative to the self-energy term, and the resulting equation is

\begin{equation}
\Sigma({k_0})=-\frac{gT}{\nu}
%EB\int \frac{d^dq}{(2\pi)^d}\frac{\omega_0^2}{(\omega_0+\omega_1\cos(q_x))^2}
\omega_0^2\left\langle\frac{1}{\omega_\mathbf{q}^2}\right\rangle\frac{1}{{-k_0} +\Sigma(k_0)}.
\end{equation}

In the dispersive case, we neglect $\omega_1(\mathbf{q})$ relative to $\omega_0$, and therefore get, for both dispersive and non-dispersive phonons

\begin{equation}
\Sigma({k_0})=-\frac{gT}{\nu}
%EB\int \frac{d^dq}{(2\pi)^d}\frac{\omega_0^2}{(\omega_0+\omega_1\cos(q_x))^2}
\frac{1}{{-k_0} +\Sigma(k_0)},
\end{equation}
which may be solved to give
\begin{equation}
\begin{split}
\Sigma(k_0) &= \frac{1}{2}\left(k_0+i\sqrt{\frac{4gT}{\nu}-k_0^2}\right)\\
G^R(\mathbf{k},k_0) &= \frac{2}{k_0-2\xi_\mathbf{k}+i\sqrt{\frac{4gT}{\nu}-k_0^2}},
\end{split}
\end{equation}
results that are correct up to order $(W/gT)^2$, due to the particle-hole symmetric spectrum. Here, $G^R(\mathbf{k},k_0)$ is the retarded Green's function at momentum $\mathbf{k}$ and frequency $k_0$. The scattering rate on the Fermi energy, $\Sigma''(0)$, is much larger than the bandwidth, and the electronic quasiparticles are not well defined.

\begin{figure}	
  \centering 
\includegraphics[width=0.5\textwidth]{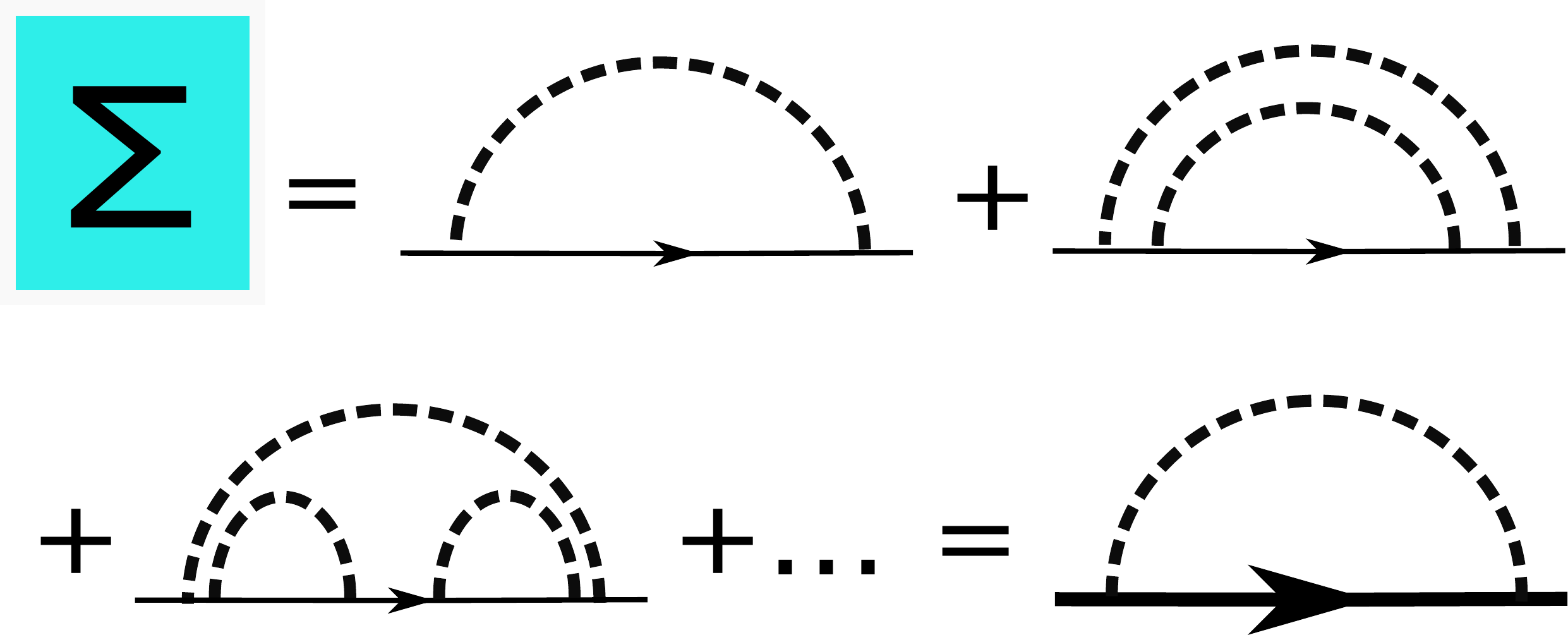}
  \caption{To lowest order in $1/N$, the full set of rainbow diagrams contributes to the electron self energy, denoted by the blue square (color online). The arrows represent bare electron propagators, dashed lines phonon propagators, and the thick arrow the fully dressed electron propagator.}
\label{fig:rainbow}
\end{figure}

It must be emphasized that, since $T\gg \omega_0$, the phonons are highly excited over the entire Brillouin zone. 
%EB; the integral over the phonon momentum $\mathbf{q}$ in Eq.~(\ref{eq:selfenergy}) is over the entire first Brillouin zone, so that the fermionic momentum $\mathbf{k+q}$ may occur outside of it. 
Therefore, the contributions of regular and umklapp scattering processes to the electronic self-energy are of the same order of magnitude. Hence, momentum is not even approximately conserved in our system.

\subsubsection{Phonon self energy}
The phonon self energy is subleading in $1/N$; however, it is of crucial importance in calculating the scrambling rate and the thermal conductivity. To leading order in $1/N$, the diagram which contributes to the self energy is shown schematically in Fig.~\ref{fig:phselfenergy}, and is given by
\begin{equation}
\Pi(\mathbf{q},i\omega_n) = -\frac{\alpha^2}{NM\omega_0}\frac{1}{\beta}\sum_{\nu_n}\int \frac{d^dk'}{(2\pi)^d}\mathcal{G}(\mathbf{k'},i\nu_n)\mathcal{G}(\mathbf{k'+q},i\nu_n+i\omega_n).
\end{equation}
Analytically continuing to real time, we get
\begin{equation}
\Pi^R(\mathbf{q},q_0) = \frac{\alpha^2}{NM\omega_0}\int\frac{d^dk'}{(2\pi)^d}\int \frac{d\epsilon}{2\pi}n_F(\epsilon)\left[A(\mathbf{k'},\epsilon)G^R(\mathbf{k'+q},\epsilon+q_0)+G^A(\mathbf{k'},\epsilon-q_0)A(\mathbf{k'+q},\epsilon)\right],
\end{equation}
where $G^{R,A}$ are retarded and advanced Green's functions, respectively, and $A(\mathbf{k},\epsilon) = -2\Im\left\{G^R(\mathbf{k},\epsilon)\right\}$ is the electron spectral function.

\begin{figure}	
  \centering 
\includegraphics[width=0.5\textwidth]{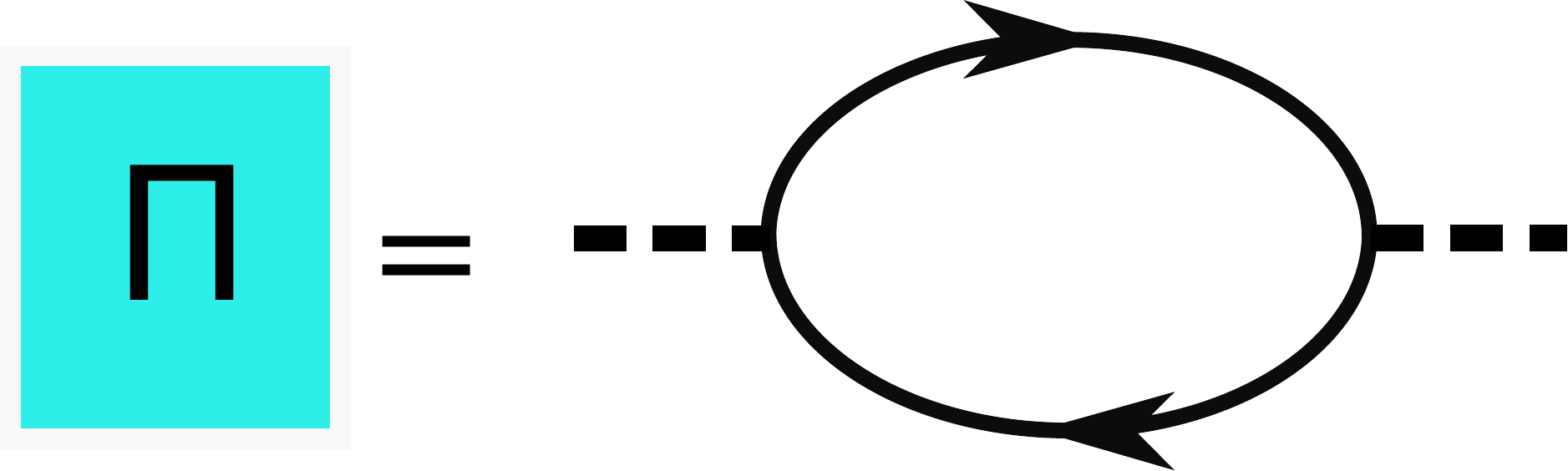}
  \caption{To lowest order in $1/N$, the only contribution to the phonon self energy is the electron bubble diagram. The arrows represent fully dressed electron propagators, while the dashed lines are bare phonon propagators.}
\label{fig:phselfenergy}
\end{figure}

The phonon lifetime is given by $\Pi''(\mathbf{q},q_0) = -\Im\Pi^R(\mathbf{q},q_0)$, which for small external frequencies $q_0\ll T$ gives

\begin{eqnarray}\label{eq:Pi}
\Pi''(\mathbf{q},q_0) &=& -\frac{\alpha^2}{2 NM\omega_0}\int \frac{d\epsilon}{2\pi}\int \frac{d^dk'}{(2\pi)^d}A(\mathbf{k'},\epsilon)A(\mathbf{k'+q},\epsilon+k_0)\left(n_F(\epsilon)-n_F(\epsilon+q_0)\right)\nonumber\\
&\approx& -q_0\frac{\alpha^2}{2 NM\omega_0}\int_{-q_0}^0 \frac{d\epsilon}{2\pi}\int\frac{d^dk'}{(2\pi)^d}A(\mathbf{k'},\epsilon)A(\mathbf{k'+q},\epsilon+q_0)[-n'_F(\epsilon)]\nonumber\\
&\approx&-\frac{1}{8\pi N}\frac{q_0\omega_0}{T}
\end{eqnarray}

As we are concerned with the spatial propagation of chaos, we will also need the velocity of the phonons. For a finite bare dispersion $\omega_1(\mathbf{q})$, the leading order contribution is simply the bare velocity $v_{ph}$. In the case of a vanishing bare velocity, $v_{ph}=0$, the leading order velocity originates in the $\mathbf{q}$ dependence of the real part of the self-energy, $\Pi'(\mathbf{q},q_0) = \Re\Pi^R(\mathbf{q},q_0)$; for $q_0=0$, this becomes to lowest order in $W/gT$

\begin{equation}
\begin{split}
\Pi'(\mathbf{q},q_0=0) &= -\frac{\alpha^2}{2 NM\omega_0}\int \frac{d\epsilon}{2\pi}n_F(\epsilon)\int \frac{d^dk}{(2\pi)^d} \\
&\times\left[A(\mathbf{k},\epsilon)\Re G^R(\mathbf{k+q},\epsilon+q_0)+\Re G^R(\mathbf{k},\epsilon-q_0)A(\mathbf{k+q},\epsilon)\right]\\
&=\frac{1}{N}\frac{\omega_0}{T}\sqrt{\frac{gT}{\nu}}\left[\frac{4\sqrt{2}}{3}-\frac{4\sqrt{2}}{15}\frac{\nu}{gT}\int \frac{d^dk}{(2\pi)^d}\xi_\mathbf{k}\xi_\mathbf{k+q}\right].
\end{split}
\end{equation}

We thus define a renormalized phonon velocity  $\tilde{v}_{ph}$ (applicable when $v_{ph}=0$):

\begin{equation}
\tilde{v}_{ph}^2 \equiv \left\langle \left(\partial_\mathbf{q}\Pi'(\mathbf{q},q_0=0) \right)^2 \right\rangle =\frac{1}{N^2}\frac{\omega_0^2}{T^2}\frac{\nu}{gT}\int \frac{d^dkd^dk'd^dq}{(2\pi)^{3d}} \xi_\mathbf{k} \xi_\mathbf{k'}v_\mathbf{k+q}v_\mathbf{k'+q},
\end{equation}

where $v_\mathbf{k} = \partial_\mathbf{k}\xi_\mathbf{k}$ is the electronic velocity.

\section{Electronic scrambling rate}
\subsection{Definition}
The scrambling rate is defined as the exponetial rate of growth of the expression
\begin{equation} \label{eq:scramb}
\left\langle\left[\hat{W}(x,t),\hat{V}(0)\right]^2\right\rangle  = \mathrm{Tr}\left\{\rho\left[\hat{W}(x,t),\hat{V}(0)\right]\left[\hat{W}(x,t),\hat{V}(0)\right]\right\}.
\end{equation}
Following Refs.~\cite{Maldacena, Stanford}, we regularize this expression by splitting the thermal density matrix ${\rho} = e^{-\beta H}/Z$ (where $Z=\mathrm{Tr}[e^{-\beta H}]$), and placing $\sqrt{\rho}$ between the two commutators in Eq.~(\ref{eq:scramb}). For electrons, we use an anti-commutator rather than a commutator~\cite{Patel}. We thus define the function
\begin{equation}
\label{eq:f}
\begin{split}
f(t) &= \frac{1}{N^2}\theta(t)\sum_{ab}\int d^dx \mathrm{Tr}\left[\sqrt{\rho}\{c_a(x,t),c^\dagger_b(0,0)\}\sqrt{\rho}\{c_a(x,t),c^\dagger_b(0,0)\}^\dagger\right]\\
&=  -\frac{1}{N^2}\theta(t)\sum_{ab}\sum_{\eta,\eta'=0,1}\int d^dx \\
&\times \frac{1}{Z}\mathrm{Tr}\left[e^{-\beta H_0}T_C\left(c^{H_0}_a(x,t)c^{\dagger,H_0}_b(0,0_\eta)c^{\dagger,H_0}_a(x,t+i\frac{\beta}{2})c^{H_0}_b(0,0_{\eta'}+i\frac{\beta}{2})e^{-i\int_CH^{H_0}_\mathrm{int}(\bar{t})d\bar{t}}\right)\right],
\end{split}
\end{equation}
where $T_C$ orders the operators along the contour $C$ which is shown in Fig.~\ref{fig:contour}, and the superscript $H_0$ denotes time evolution with respect to $H_0$. The subscript $t_\eta$ with $\eta = 0,1$ refers to the forward/backward branch of each fold, respectively, and $c(t+i\tau) = e^{-H_0\tau}c^{H_0}(t)e^{H_0\tau}$. [Note that $e^{\beta H_0/2}e^{-\beta H_0/2} = 1$ has been inserted in Eq.~(\ref{eq:f}).] This form is amenable to perturbation theory, as Wick's theorem may be used.

\begin{figure}
\label{fig:contour}
  \centering
    \includegraphics[width=0.5\textwidth]{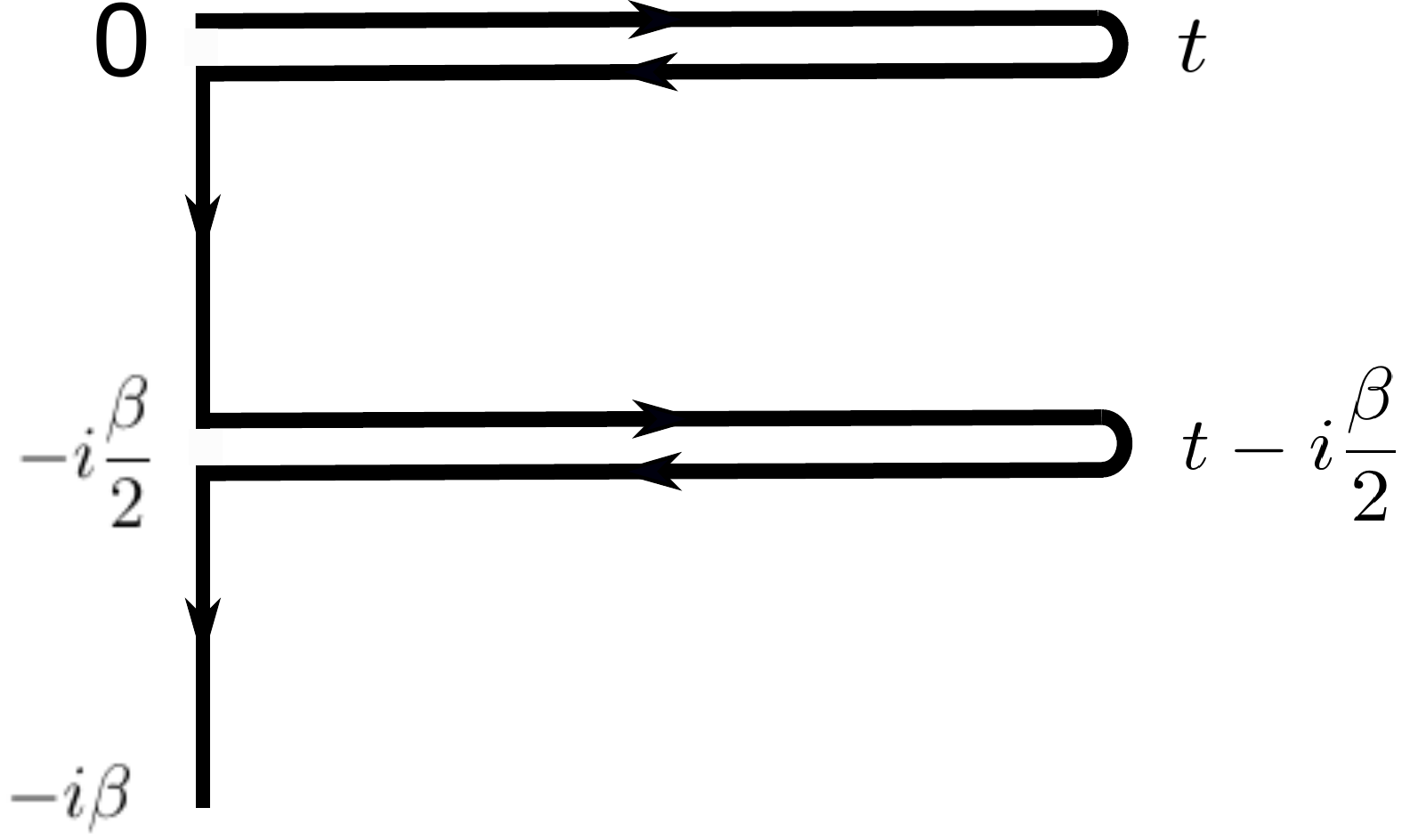}
  \caption{The contour in the complex time plane.}
\end{figure}

We expand the exponential $e^{-i\int_CH^{H_0}_\mathrm{int}(\bar{t})d\bar{t}}$, and organize the series in powers of $1/N$. For example, the $\alpha^2$ term is

\begin{equation}
\begin{split}
& \frac{1}{N^2}\theta(t)\sum_{ab}\sum_{\eta,\eta'}\int d^dx \\
&\times \mathrm{Tr}\left[e^{-\beta H_0}T_C\left(c^{H_0}_a(x,t)c^{\dagger,H_0}_b(0,0_\eta)c^{\dagger,H_0}_a(x,t+i\frac{\beta}{2})c^{H_0}_b(0,0_{\eta'}+i\frac{\beta}{2})\right.\right.\\
&\left.\left.\int_C d\bar{t}_1d\bar{t}_2H^{H_0}_\mathrm{int}(\bar{t}_1)H^{H_0}_\mathrm{int}(\bar{t}_2)\right)\right]
\end{split}
\end{equation}

The form of the contour $C$ leads to several qualitatively different contributions:
\begin{enumerate}
\item{$\bar{t}_1$ and $\bar{t}_2$ are both on the same real time fold: this leads to a self-energy correction to the propagators, which are taken into account by using the self-energies calculated in the previous section.}

\item{$\bar{t}_1$ and $\bar{t}_2$ are both imaginary times: this leads to corrections to the thermal state, which is also taken into account by considering the fully dressed propagators. Contractions between the external operators $c_a(x,t)$, $c_b(0)$ and vertex operators at imaginary times vanish due to the summation over $\eta,\eta'$.}

\item{$\bar{t}_1$ and $\bar{t}_2$ are on different real time folds: this leads to a qualitatively new contribution, with contractions of the form (where flavor and spatial indices have been suppressed),
\begin{equation}
\mathrm{Tr}\left[e^{-\beta H_0} X(t)X(t'+i\beta/2)\right]\mbox{,	} \mathrm{Tr}\left[e^{-\beta H_0} c(t)c(t'+i\beta/2)\right]
\end{equation}
and must be taken into account.}

\item{$\bar{t}_1$ is real and $\bar{t}_2$ is imaginary: this contribution is subleading in $1/N$.}
\end{enumerate}

\subsection{Bethe-Salpeter equation for $f(\omega)$}
We obtain $f(\omega)$ by summing an infinite series of diagrams. The summation is equivalent to solving a self-consistent Bethe-Salpeter equation for the two-particle correlator. To leading order in $1/N$, the Bethe-Salpeter equation, which is schematically shown in Fig.~\ref{fig:elchaos}, is given by

\begin{figure}
  \centering
    \includegraphics[width=0.9\textwidth]{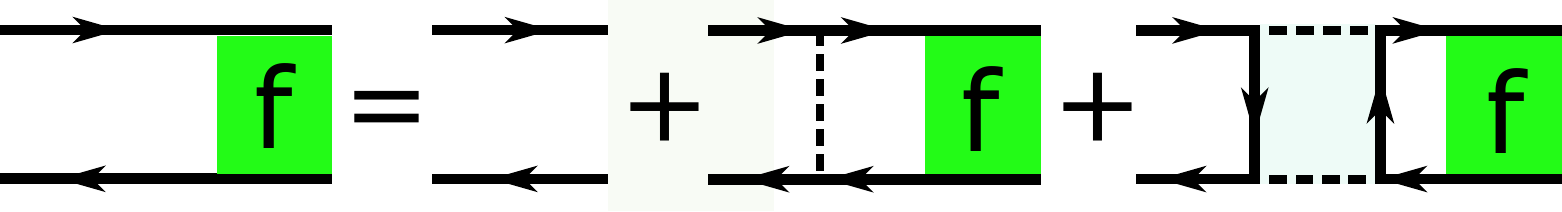}
  \caption{The Bethe-Salpeter equation for $f(\omega)$. Horizontal lines represent retarded/advanced propagators, with the top line living at the upper branch and the bottom line on the lower branch $-i\beta/2$. Vertical lines are Wightman propagators, in this case the dashed line represents a phonon. Only the single-phonon Wightman diagram enters to leading order in $1/N$.}\label{fig:elchaos}
\end{figure}

\begin{eqnarray}\label{eq:Bethe}
f(z) &=& \int_0^\infty dt\, e^{izt} \tilde{f}(t)\\
f(\omega) &=& \frac{1}{N}\int \frac{d^{d+1}k}{(2\pi)^{d+1}} f(\mathbf{k},k_0,\omega)\nonumber\\
f(\mathbf{k},k_0,\omega) &=& G^R(\mathbf{k},k_0)G^A(\mathbf{k},k_0-\omega)\nonumber\\
&\times&\left[1-\alpha^2\int \frac{d^{d+1}k'}{(2\pi)^{d+1}} D^W(\mathbf{k-k'},k_0-k'_0)f(\mathbf{k'},k'_0,\omega)\right.\nonumber\\
&+&\left.\frac{1}{N}\alpha^4\int \frac{d^{d+1}k'}{(2\pi)^{d+1}} \mathcal{F}(k,k',\omega)f(\mathbf{k'},k'_0,\omega)
\right]\nonumber\\
&+&O\left(\frac{1}{N}\right).\nonumber
\end{eqnarray}

Here,
\begin{eqnarray}\label{Eq.F}
\mathcal{F}(k,k',\omega) &=& \int \frac{d^{d+1}k_1}{(2\pi)^{d+1}}  D^R(\mathbf{k_1},k_{10})D^A(\mathbf{k_1},k_{10}-\omega)\nonumber\\
&&\times G^W(\mathbf{k-k_1},k_0-k_{10})G^W(\mathbf{k'-k_1},k'_0-k_{10})
\end{eqnarray}
and we define
\begin{eqnarray}
D^W(x_1-x_2,t_1-t_2) \equiv \frac{1}{Z}\mathrm{Tr}\left[e^{-\beta H_0} X_{ab}(x_1,t_{1\eta})X_{ab}(x_2,t_{2\eta'}+i\frac{\beta}{2}) \right] 
\end{eqnarray}
as the phonon Wightman propagator, and in the same way 
\begin{eqnarray}
G^W(x_1-x_2,t_1-t_2) \equiv \frac{1}{Z}\mathrm{Tr}\left[e^{-\beta H_0} c_{a}(x_1,t_{1\eta})c^\dagger_{a}(x_2,t_{2\eta'}+i\frac{\beta}{2}) \right].
\end{eqnarray}
An explicit calculation shows that (see, e.g., Ref.~\cite{Patel})
\begin{eqnarray}\label{Wightman}
D^W(\mathbf{q},q_0) &=& \frac{B(\mathbf{q},q_0)}{2\sinh(q_0/2T)} = \frac{2\pi}{2M\omega_0\sinh(\omega_0/2T)}\left[\delta(q_0-\omega_0)+\delta(q_0+\omega_0)\right]\nonumber \\
G^W(\mathbf{k},k_0) &=& \frac{A(\mathbf{k},k_0)}{2\cosh(k_0/2T)},
\end{eqnarray}
with $A$, $B$ the electron and phonon spectral functions, respectively.

Naively, the $\alpha^4$ in Eq.~(\ref{eq:Bethe}) appears to be sub-leading in powers of $1/N$. However, as we shall show below, it is actually of the same order as the $\alpha^2$ term, and must be kept. This is because this term diverges unless the self-energy of the phonons, which is of order $1/N$, is used.

If $f(t)$ has an exponentially growing part, $f(t) = \frac{1}{N}e^{\lambda_L t}+O\left(\frac{1}{N^2}\right)$, its Laplace transform should divrege for $z = i\lambda_L$. We look for divergences in $f(\mathbf{k},k_0,\omega=i\lambda_L)$ and for this imaginary frequency we may ignore the inhomogenous term in Eq.~(\ref{eq:Bethe}) to get
\begin{eqnarray}\label{eq:bethe}
f(\mathbf{k},k_0,\omega) &=& G^R(\mathbf{k},k_0)G^A(\mathbf{k},k_0-\omega) \left[-\alpha^2\int \frac{d^{d+1}k'}{(2\pi)^{d+1}} D^W(\mathbf{k-k'},k_0-k'_0)f(\mathbf{k'},k'_0,\omega)\right.\nonumber\\
&+&\left.\frac{1}{N}\alpha^4\int \frac{d^{d+1}k'}{(2\pi)^{d+1}} \mathcal{F}(k,k',\omega)f(\mathbf{k'},k'_0,\omega)\right].
\end{eqnarray}
At long times, $f(t)$ is expected to saturate. This effect is captured by higher order terms in $1/N$~\cite{Stanford}.

\subsection{Solution of the Bethe-Salpeter equation}

In the high temperature regime, the Green's functions are given by the results of section \ref{selfenergies}. The term $G^R(\mathbf{k},k_0) G^A(\mathbf{k},k_0-\omega)$ may be approximated, for $k_0,\omega\ll \sqrt{gT/\nu}$, by
\begin{eqnarray}
G^R(\mathbf{k},k_0) G^A(\mathbf{k},k_0-\omega) \approx \frac{\nu}{2gT}.
\end{eqnarray}

Since the phonon propagators are unrenormalized (even at high temperature) to leading order in $1/N$, they are strongly peaked on shell. 
%EBIn the following secction, the 
The term $D^R(\mathbf{k},k_{0})D^A(\mathbf{k},k_{0}-\omega)$, that appears in Eq.~(\ref{Eq.F}) above,   %will be encountered; it 
will therefore be replaced by: %EB $\delta$-functions times residues at the poles, i.e.
\begin{eqnarray}\label{Eq:phprop}
%&&D^R(\mathbf{k},k_{0})D^A(\mathbf{k},k_{0}-\omega)\\
 D^R(\mathbf{k},k_{0})D^A(\mathbf{k},k_{0}-\omega) &=& \frac{1}{M^2\omega_0^2}\left(\frac{1}{k_0-\omega_0+i\Pi''(k_0)}-\frac{1}{k_0+\omega_0+i\Pi''(k_0)}\right)\nonumber\\
&&\times\left(\frac{1}{k_0-\omega-\omega_0-i\Pi''(k_0-\omega)}-\frac{1}{k_0-\omega+\omega_0-i\Pi''(k_0-\omega)}\right)\nonumber\\
 &\approx&\frac{1}{M^2\omega_0^2}
\frac{-2\pi i}{\omega+2i\Pi''(k_0)}\left(\delta(k_{0}-\omega_0)+\delta(k_{0}+\omega_0)\right).
\end{eqnarray}
In the last line, we kept only terms that diverge at $\omega=0$ in the limit $1/N\rightarrow 0$. Note that, by Eq.~(\ref{eq:Pi}), $\Pi''(k_0)\propto 1/N$. Therefore, the $\alpha^4$ term in Eq.~(\ref{eq:Bethe}) is in fact of the same order in $1/N$ as the $\alpha^2$ term.

The Bethe-Salpeter equation is therefore (assuming that $\omega$ is much smaller than the electronic energy scale $gT/\nu$)

\begin{eqnarray}
f(k_0,\omega) &=& \frac{\nu}{2gT}\left[-2\alpha^2\frac{T}{K}f(k_0,\omega) + \frac{-i}{4N}\frac{\alpha^4}{M^2\omega_0^2}
\int \frac{dk'_0}{2\pi}\int \frac{dk_{10}}{2\pi}\frac{\delta(k_{10}-\omega_0)+\delta(k_{10}+\omega_0)}{\omega+2i\Pi''(k_{10})}\right.\nonumber\\
&\times&\left.\frac{A(k_0-k_{10})}{\cosh((k_0-k_{10})/2T)}\frac{A(k'_0-k_{10})}{\cosh((k'_0-k_{10})/2T)}f(k'_0,\omega)\right]\nonumber\\
\end{eqnarray}
using the relation $g = \alpha^2 \nu/K$, and setting the argument of both the electron spectral functions to zero (as $T\ll \sqrt{gT/\nu}$), this becomes
\begin{eqnarray}
2f(k_0,\omega) &=& \frac{-i}{4N}\frac{\nu}{2gT}\frac{\alpha^4}{M^2\omega_0^2}
\int \frac{dk'_0}{2\pi}\int \frac{dk_{10}}{2\pi}\frac{\delta(k_{10}-\omega_0)+\delta(k_{10}+\omega_0)}{\omega+2i\Gamma(k_{10})}\nonumber\\
&\times&\frac{A(0)}{\cosh((k_0-k_{10})/2T)}\frac{A(0)}{\cosh((k'_0-k_{10})/2T)}f(k'_0,\omega)\\
&=& \frac{-i}{4N}\frac{\omega_0^2}{T^2}
\left[\frac{1}{\omega+2i\Gamma}+\frac{1}{\omega-2i\Gamma}\right]
\frac{1}{\cosh(k_0/2T)}\int \frac{dk'_0}{2\pi}\frac{1}{\cosh(k'_0/2T)}f(k'_0,\omega)\nonumber,
\end{eqnarray}
where in the last line we have neglected $\omega_0$ relative to the electronic energy scales, and we define
\begin{eqnarray}
\Gamma = \Pi''(\omega_0) = -\frac{1}{2\pi N}\frac{\omega_0^2}{T}.
\end{eqnarray}
We make the ansatz $f(k_0,\omega) = \tilde{f}(\omega)/\cosh(k_0/2T)$ and perform the integral over $k'_0$ to get
\begin{eqnarray}
2\tilde{f}(\omega)= i\Gamma
\left[\frac{1}{\omega+2i\Gamma}+\frac{1}{\omega-2i\Gamma}\right]
\tilde{f}(\omega)
\end{eqnarray}
which is solved by
\begin{eqnarray}
\omega = \frac{1}{2}i|\Gamma|\left(\sqrt{17}-1\right)
\end{eqnarray}

We therefore conclude that the electronic scrambling rate is
\begin{eqnarray}
\lambda_L \approx 0.25\frac{1}{N}\frac{\omega_0^2}{T}.
\end{eqnarray}

\subsection{Butterfly velocity and diffusion constant}
The out-of-time-ordered correlation function, evaluated at two distinct, spatially separated points, allows us to quantify the rate of propagation of scrambling in both space and time. More specifically, we define the function

\begin{eqnarray}
f(x,t) &=& \frac{1}{N^2}\theta(t)\sum_{ab}\mathrm{Tr}\left[\sqrt{\rho}\{c_a(x,t),c^\dagger_b(0,0)\}\sqrt{\rho}\{c_a(x,t),c^\dagger_b(0,0)\}^\dagger\right],\nonumber\\
\end{eqnarray}
which is expected to obey
\begin{equation}
f(x,t)\sim e^{\lambda_L(t-|x|/v)-x^2/D_Lt}.
\end{equation}

We study its Fourier transform, $f(\mathbf{q},t) = \int d^dx e^{-i\mathbf{q}\mathbf{x}} f(x,t)$, which has two limiting behaviors:
\begin{enumerate}
\item{If the ballistic term is absent, $v=0$, we get 
\begin{equation}
f(q,t) \sim e^{(\lambda_L-\frac{1}{4}D_Lq^2)t},
\end{equation}
so that the velocity scale for chaos propagation, which is the butterfly velocity, is given by $\tilde{v} = \sqrt{D_L\lambda_L}$, and $D_L$ represents the leading correction to the Lyapunov exponent at small $q$:
\begin{equation}
\lambda_L(q) = \lambda_L(q=0)-\frac{1}{4}D_L q^2  + O(q^4).
\end{equation}
}
\item{If the diffusive term is absent, we get
\begin{equation}
f(q,t) \sim \frac{v\lambda_L}{\lambda_L^2+v^2q^2}e^{\lambda_Lt},
\end{equation}
and in this limit the butterfly velocity is given by $v$, which characterizes the dependence of the 
%EBchange to the amplitude of 
prefactor of the exponent in $f(q,t)$ on $q$, rather than the $q$ dependence of the scrambling rate.}
\end{enumerate}
We therefore define the butterfly velocity as by $v_B = \max\left(v,\tilde{v}\right)$, where $v^2$ is the coefficient of the $q^2$ term in the prefactor of the exponent in $f(q,t)$ and $\tilde{v}$ is the coefficient of the $q^2$ term in $\lambda_L(q)$. More precisely,
\begin{equation}\label{eq:v_B}
	v_B^2 =\lim_{q\rightarrow 0}\frac{\lambda_L^2}{q^2} \frac{f(q,t=\lambda_L^{-1})-f(t=\lambda_L^{-1})}{f(t=\lambda_L^{-1})}.
\end{equation}

The $q$-dependent Bethe-Salpeter equation for the Fourier transform $f(\mathbf{q},\omega)$ is given by

\begin{eqnarray}\label{eq:Bethe2_sup}
f(\mathbf{q},\omega) &=& \frac{1}{N}\int\frac{d^{d+1}k}{(2\pi)^{d+1}} f(\mathbf{k},k_0,\mathbf{q},\omega)\\
f(\mathbf{k},k_0,\mathbf{q},\omega) &=& G^R(\mathbf{k+q},k_0)G^A(\mathbf{k},k_0-\omega)\nonumber\\
&\times&\left[1-\alpha^2\int \frac{d^{d+1}k'}{(2\pi)^{d+1}} D^W(\mathbf{k-k'},k_0-k'_0)f(\mathbf{k'},k_0',\mathbf{q},\omega)\right.\nonumber\\
&+&\left.\frac{1}{N}\alpha^4\int \frac{d^{d+1}k'}{(2\pi)^{d+1}} \mathcal{F}(k,k',\mathbf{q},\omega)f(\mathbf{k'},k_0',\mathbf{q},\omega)
\right]\nonumber\\
&+&O\left(\frac{1}{N}\right).\nonumber
\end{eqnarray}

Here
\begin{eqnarray}
\mathcal{F}(k,k',\mathbf{q},\omega) &=& \int \frac{d^{d+1}k_1}{(2\pi)^{d+1}} D^R(\mathbf{k_1+q},k_{10})D^A(\mathbf{k_1},k_{10}-\omega)\nonumber\\
&&\times G^W(\mathbf{k-k_1},k_0-k_{10})G^W(\mathbf{k'-k_1},k'_0-k_{10}).
\label{Eq.Fq}
\end{eqnarray}

{This Bethe-Salpeter equation can be schematically written in a matrix form %EB(replacing the integral with matrix multiplication)
\begin{equation}
f(k,q,\omega) = f_0(k,q,\omega) + \int_{k'} \mathcal{M}(k,k',q,\omega) f(k',q,\omega),
\end{equation}
understood as an equation for the vector $f(k,q,\omega)$ (indexed by $k$), with an inhomogenous term $f_0(k,q,\omega)$. $\mathcal{M}$ is an operator that can be read off from Eq.~(\ref{eq:Bethe2_sup}).} Finding the Lyapunov exponent corresponds to finding the $\omega$ for which $\mathcal{M}$ has an eigenvalue $\lambda = 1$, and for frequencies for which $1-\mathcal{M}(k,q,\omega)$ has an inverse, $f(k,q,\omega)$ is given by
{
\begin{equation}
f(k,q,\omega) = \int_{k'} \left[1-\mathcal{M}(k,k',q,\omega)\right]^{-1} f_0(k',q,\omega)
\end{equation}
}
Therefore, the $q$-dependence of $\lambda_L$ originates in the $q$-dependence of $\mathcal{M}$, while the $q$-dependence of the amplitude of $f(k,q,\omega)$ has contributions from the $q$-dependence of both $\mathcal{M}$ and $f_0$.

Any $q$-dependence originates from the electron or phonon propagator pairs, which we expand for small $\mathbf{q}$ to give (for small $k_0$)
\begin{equation}
\begin{split}
G^R(\mathbf{k+q},k_0)G^A(\mathbf{k},k_0-\omega) &= G^R(\mathbf{k},k_0)G^A(\mathbf{k},k_0-\omega) + \left(v_\mathbf{k}\mathbf{q}\right)^2G^R(\mathbf{k},k_0)^3G^A(\mathbf{k},k_0-\omega)\\
&\approx\frac{\nu}{2gT} - \left(\frac{\nu}{2gT}\right)^2 v_\mathbf{k}^2q^2\\
D^R(\mathbf{k+q},k_0)D^A(\mathbf{k},k_0-\omega)&\approx \frac{1}{M^2\omega_0^2}
\left(\delta(k_{0}-\omega_0)+\delta(k_{0}+\omega_0)\right)\\
&\times \frac{-2\pi i}{\omega+\omega_\mathbf{k+q}-\omega_\mathbf{k}+\Pi'(\mathbf{k+q})-\Pi'(\mathbf{k})+2i\Pi''(k_0)}\\
&\approx \frac{-2\pi i}{M^2\omega_0^2}
\left(\delta(k_{0}-\omega_0)+\delta(k_{0}+\omega_0)\right)\\
&\times\left[\frac{1}{\omega+2i\Pi''(k_0)}-\frac{\left(\partial_{\mathbf{k}}\left[\omega_\mathbf{k}+\Pi'(\mathbf{k})\right]\right)^2q^2}{\left(\omega+2i\Pi''(k_0)\right)^3}\right]
\end{split}
\end{equation}
(the terms linear in $v_\mathbf{k}, \partial_{\mathbf{k}}\Pi'(\mathbf{k})$ will vanish under integration.) 

The dominant contribution to the $q$ dependence %EBshift 
of the integral operator $\mathcal{M}(k,k',q,\omega)$ comes from the $D^R(\mathbf{k+q},k_0)D^A(\mathbf{k},k_0-\omega)$ term. We roughly estimate the magnitude of the leading $q$ dependence of $\mathcal{M}$ as
\begin{equation} \label{eq:deltaM}
 \frac{| \delta \mathcal{M} |}{ |\mathcal{M}|} \sim q^2\frac{1}{\Gamma^2}\int \frac{d^dk}{(2\pi)^d} \left(\partial_{\mathbf{k}}\left[\omega_\mathbf{k}+\Pi'(\mathbf{k})\right]\right)^2.
\end{equation}

The $q$-dependence of the inhomogeneous term, $\int d^dk G^R(\mathbf{k+q},k_0)G^A(\mathbf{k},k_0-\omega)$, is smaller by {at least} a factor of $\omega_0^2/E_F^2 \ll 1$,  %EBin the case of phonons without a bare dispersion, and we therefore neglect it:
\begin{equation}
\frac{|\delta f_0|}{|f_0|} \sim q^2 \frac{\nu}{gT} \int \frac{d^dk}{(2\pi)^d} \left(\partial_{\mathbf{k}}\xi_\mathbf{k}\right)^2.
\end{equation}

Therefore, using Eqs.~(\ref{eq:deltaM}) and (\ref{eq:v_B}), we conclude that

\begin{equation}
v_B^2 \sim  \left \{
  \begin{tabular}{cc}
  $v_{ph}^2$ & (Dispersive phonons)   \\
  $\tilde{v}_{ph}^2$ & (Non-dispersive phonons).   
  \end{tabular}
\right.
\end{equation}

\section{Phonon scrambling rate}
\subsection{Definition}
{We now calculate the scrambling rate and corresponding butterfly velocity of defined in terms of the phonon coordinates. The calculation follows similar lines to that of the electronic scrambling rate.} 

We define the function
\begin{eqnarray}
h(t) = \frac{1}{N^4}\theta(t)\sum_{abcd}\int d^dx \mathrm{Tr}\left[\sqrt{\rho}\left[X_{ab}(x,t),X_{cd}(0,0)\right]\sqrt{\rho}\left[X_{ab}(x,t),X_{cd}(0,0)\right]^\dagger\right],\nonumber\\
\end{eqnarray}
and its Laplace transform $h(\omega)$.
\subsection{Bethe-Salpeter equation}
The Bethe-Salpeter equation for $h(\omega)$, represented schematically in Fig.~\ref{fig:phchaos}, is 
\begin{eqnarray}
h(\omega) &=& \frac{1}{N^2}\int \frac{d^{d+1}k}{(2\pi)^{d+1}} h(\mathbf{k},k_0,\omega)\\
h(\mathbf{k},k_0,\omega)  &=& D^R(\mathbf{k},k_0)D^A(\mathbf{k},k_0-\omega)\nonumber\\
&\times&\left[1+\frac{\alpha^4}{N}\frac{d^{d+1}k'}{(2\pi)^{d+1}}\mathcal{H}(k,k',\omega)h(\mathbf{k'},k_0',\omega)+O\left(\frac{1}{N^2}\right) \right]\nonumber
\end{eqnarray}
with
\begin{eqnarray}
\mathcal{H}(k,k',\omega) = \int \frac{d^{d+1}k_1}{(2\pi)^{d+1}} G^R(\mathbf{k_1},k_{10})G^A(\mathbf{k_1},k_{10}-\omega)G^W(\mathbf{k-k_1},k_0-k_{10})G^W(\mathbf{k'-k_1},k'_0-k_{10}).\nonumber\\
\end{eqnarray}
%EBas shown schematically in Fig.\ref{fig:phchaos}. 

\begin{figure}
  \centering
    \includegraphics[width=0.5\textwidth]{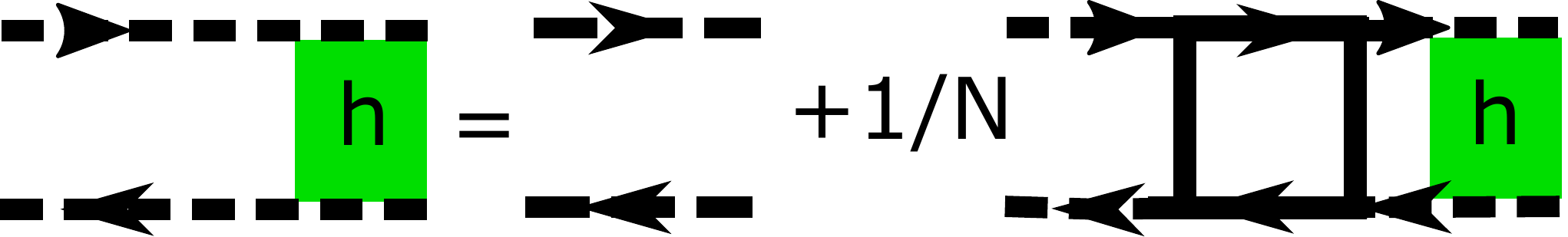}
  \caption{The Bethe-Salpeter equation for $h(\omega)$. Horizontal lines represent retarded/advanced propagators, with the top line living at the upper branch and the bottom line on the lower branch $-i\beta/2$. Vertical lines are Wightman propagators, in this case the dashed lines represent phonons and full lines electrons. This is the only diagram to leading order in $1/N$.}
\label{fig:phchaos}
\end{figure}

As before, to find the scrambling rate, we ignore the homogenous term. This gives the equation
\begin{eqnarray}
h(\mathbf{k},k_0,\omega)  &=& \frac{\alpha^4}{N}D^R(\mathbf{k},k_0)D^A(\mathbf{k},k_0-\omega)\int \frac{d^{d+1}k'}{(2\pi)^{d+1}}\mathcal{H}(k,k',\omega)h(\mathbf{k'},k_0',\omega).
\end{eqnarray}

As was discussed around Eq.~(\ref{Eq:phprop}) above, the phonon propagators are strongly peaked, and we replace them with $\delta$-functions and residues:

\begin{eqnarray}
h(\mathbf{k},k_0,\omega)  = \frac{-i}{N}\frac{\alpha^4}{M^2\omega_0^2}
\frac{2\pi}{\omega+2i\Gamma(k_0)}\left(\delta(k_{0}-\omega_0)+\delta(k_{0}+\omega_0)\right)
\int \frac{d^{d+1}k'}{(2\pi)^{d+1}}\mathcal{H}(k,k',\omega)h(\mathbf{k'},k_0',\omega)\nonumber\\
\end{eqnarray}

At high temperatures, we set $\omega_0=0$ in all electronic quantities, and get
\begin{eqnarray}
h(\mathbf{k},k_0,\omega)  &=& \frac{-i}{N}\frac{\alpha^4}{M^2\omega_0^2}
{2\pi}\delta(k_{0})\left[\frac{1}{\omega+2i\Gamma}+\frac{1}{\omega-2i\Gamma}\right]\\
&\times&\int \frac{d^{d+1}k'}{(2\pi)^{d+1}}\frac{d^{d+1}k_1}{(2\pi)^{d+1}} G^R(\mathbf{k_1},k_{10})G^A(\mathbf{k_1},k_{10}-\omega)\nonumber\\
&&\times G^W(\mathbf{k-k_1},k_0-k_{10})G^W(\mathbf{k'-k_1},k'_0-k_{10})h(\mathbf{k'},k_0',\omega).\nonumber
\end{eqnarray}

We make the ansatz $h(\mathbf{k},k_0,\omega) = \tilde{h}(\omega)2\pi\delta(k_0)$ and get

\begin{eqnarray}
\tilde{h}(\omega) &=& \frac{-i}{N}\frac{\alpha^4}{M^2\omega_0^2}\left[\frac{1}{\omega+2i\Gamma}+\frac{1}{\omega-2i\Gamma}\right]\\
&\times&\int \frac{d^dk'}{(2\pi)^d} \frac{d^{d+1}k_1}{(2\pi)^{d+1}}G^R(\mathbf{k_1},k_{10})G^A(\mathbf{k_1},k_{10}-\omega)\nonumber\\
&&\times G^W(\mathbf{k-k_1},k_0-k_{10})G^W(\mathbf{k'-k_1},-k_{10})\tilde{h}(\omega)\nonumber\\
\end{eqnarray}

Inserting in this expression the high-$T$ form of the electron propagators, we get (all the electron propagators have $\epsilon=0$ for their argument, as $T\ll E_F$) 

\begin{eqnarray}
1 &=& \frac{-i}{4N}\frac{\omega_0^2}{T^2}\left[\frac{1}{\omega+2i\Gamma}+\frac{1}{\omega-2i\Gamma}\right]
\int\frac{dk_{10}}{2\pi}\frac{1}{\cosh^2(k_{10})}\\
&=& i\Gamma\left[\frac{1}{\omega+2i\Gamma}+\frac{1}{\omega-2i\Gamma}\right],\nonumber
\end{eqnarray}
which up to a factor of two gives the same scrambling rate as the electrons.

In the same way, the butterfly velocity may be calculated, and it results in the same velocity as for the electrons, {up to a numerical prefactor.}
\section{Thermal conductivity}

\subsection{Dispersive phonons}
In the case of dispersive phonons, the thermal conductivity is dominated by the contribution of the $N^2$ long lived phonons. These phonons are well defined quasiparticles, and therefore the thermal conductivity may be calculated using the semiclassical result:
\begin{equation}
\kappa = N^2 \frac{v_{ph}^2}{\Gamma} = N^3 v_{ph}^2\frac{T}{\omega_0^2}
\end{equation}

The contribution of the electrons and the interaction terms are calculated in the next section, and are shown to be subleading in $1/N$.

\subsection{Dispersionless phonons}
\subsubsection{Thermal current operator}
In this case, we use the Kubo formula to calculate the thermal conductivity, which requires the definition of the thermal current operator. For simplicity, we will calculate the thermal conductivity in a one-dimensional version of the system; the extension to higher dimensions is straightforward. 
The Hamiltonian is given by

\begin{eqnarray}
H &=& \sum_iH_i = \sum_i\left[H^{el}_i+H^{ph}_i+H^{int}_i\right]
\end{eqnarray}
where
\begin{eqnarray}
H^{el}_i &=& -\frac{1}{2}t\sum_a\left[c^\dagger_{a,i}c_{a,i+1}+c^\dagger_{a,i+1}c_{a,i}+c^\dagger_{a,i}c_{a,i-1}+c^\dagger_{a,i-1}c_{a,i}\right]\nonumber\\
H^{ph}_i&=&\sum_{ab}\left[\frac{P^2_{ab,i}}{2M}+\frac{1}{2}M\omega_0^2X^2_{ab,i}\right]\nonumber\\
H^{int}_i&=&\frac{\alpha}{\sqrt{N}}\sum_{ab}X_{ab,i}c^\dagger_{a,i}c_{b,i}
\end{eqnarray}

The time derivative of the single-site energy density is given by

\begin{eqnarray}\label{eq:commutators}
\frac{d}{dt}H_i = i[H,H_i] = i[H^{el},H^{el}_i] + i[H^{int},H^{el}_i] + i[H^{el},H_i^{int}]
\end{eqnarray}
where the rest of the terms vanish because of the locality of $H^{ph}_i$ and $H^{int}_i$. Here, $H^{el} = 
\sum_i H^{el}_i$, and similarly for the phonon and interaction terms. The first term on the right hand side results in the usual electronic thermal current, which is
\begin{eqnarray}
J^{el}_E = \sum_a\int \frac{d^dk}{(2\pi)^d} v_\mathbf{k}\xi_\mathbf{k}c^\dagger_a(\mathbf{k})c_a(\mathbf{k})
\end{eqnarray}

The other two terms of Eq.(\ref{eq:commutators}) result in the following contribution, which we term the phonon assisted (pa) contribution:
\begin{eqnarray}
\frac{d}{dt}H^{pa}_i  &=&  i[H^{int},H^{el}_i] + i[H^{el},H_i^{int}]\\
&=&-\frac{1}{2}\frac{i\alpha}{\sqrt{N}}\sum_{ab}\left[\left(X_{ab,i}-X_{ab,i-1}\right)\left(c^\dagger_{a,i-1}c_{b,i}-c^\dagger_{a,i}c_{b,i-1}\right)\right.\nonumber\\
&&-\left.\left(X_{ab,i}+X_{ab,i+1}\right)\left(c^\dagger_{a,i}c_{b,i+1}-c^\dagger_{a,i+1}c_{b,i}\right)\right]\nonumber
\end{eqnarray}

which corresponds to the thermal current (see, e.g., \cite{Mahan})
\begin{eqnarray}
J^{pa}_E &=& \sum_i l_i\frac{d}{dt}H^{pa}_i\\
&=&\frac{\alpha }{\sqrt{N}}\sum_{ab}\int \frac{d^dk d^dq}{(2\pi)^{2d}}\left[v_{\mathbf{k}}+v_{\mathbf{k'}}\right]X_{ab}(\mathbf{k-k'})c^{\dagger}_{a}(\mathbf{k})c_{b}(\mathbf{k'})\nonumber
\end{eqnarray}
where $l_i$ is the position of site $i$.

Therefore, we get two contributions to the thermal current:
\begin{eqnarray}
J_E = J^{el}_E+J^{pa}_E,
\end{eqnarray}

which can be shown to be equal to

\begin{equation}
J_E = \frac{1}{2}i\sum_a\int \frac{d^dk}{(2\pi)^d}v_\mathbf{k}\left[\dot{c}^\dagger_a(\mathbf{k})c_a(\mathbf{k})-c^\dagger_a(\mathbf{k})\dot{c}_a(\mathbf{k})\right]
\end{equation}

with $\dot{c}_a(\mathbf{k}) = i[H,c_a(\mathbf{k})]$.

\subsubsection{Kubo formula}
We calculate the thermal conductivity using the Luttinger prescription\cite{Luttinger1,Luttinger2,Shastry}, which states that
\begin{equation}
\kappa = -\frac{1}{T}\lim_{\omega\rightarrow 0}\frac{\Im\left[\Pi^{J_QJ_Q}(\omega)\right]}{\omega},
\end{equation}
where $\Pi^{J_QJ_Q}(\omega)$ is the retarded thermal-current thermal-current correlation function. 

\subsubsection{Naive leading order in $1/N$}

Expanding in powers of $1/N$, there is a single diagram which contributes to lowest order in $1/N$; it is shown in Fig. \ref{fig:diagrams1}. Vertex corrections appear only in subleading orders in $\omega_1/\omega_0$; they all vanish if the phonon propagator is momentum independent.

\begin{figure}	
  \centering 
\includegraphics[width=0.5\textwidth]{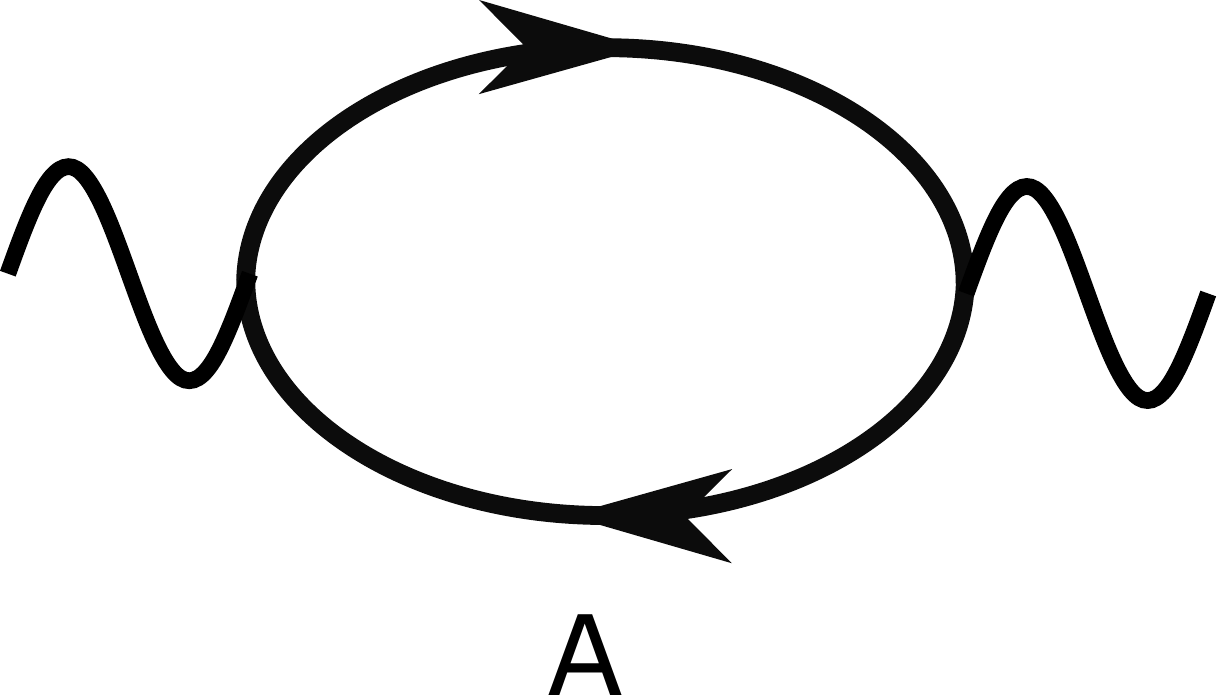}
  \caption{The diagram which contributes to the thermal conductivity at leading order in $1/N$ and in $\omega_1/\omega_0$. Black arrows are fully dressed fermionic propagators, dashed lines are phonon propagators, and the squiggly line represent $J_E$, which comes with a time derivative.}
\label{fig:diagrams1}
\end{figure}

Diagram $A$ is given by (using the Matsubara electron Green's function, $G_\mathbf{k}(i\nu_n)$, and exchanging the time derivative to an additional factor of $\nu_n$)
\begin{equation}
\begin{split}
\Pi_A(i\omega_n) &=\frac{N}{\beta}\sum_{\nu_n}\nu_n^2\int \frac{d^dk}{(2\pi)^d} v^2_\mathbf{k}G_\mathbf{k}(i\nu_n)G_\mathbf{k}(i\nu_n+i\omega_n)
\end{split}
\end{equation}

Performing the Matsubara summation and analytically continuing $i\omega_n\rightarrow \omega+i\delta$, we get
\begin{equation}
\Pi_A(\omega)=N \int \frac{d^dk}{(2\pi)^d} v^2_\mathbf{k}\int \frac{d\epsilon}{2\pi} \left[n_F(\epsilon)\epsilon^2 A(\mathbf{k},\epsilon)G^R_\mathbf{k}(\epsilon+\omega)+n_F(\epsilon)(\epsilon-\omega)^2 A(\mathbf{k},\epsilon)G^A_\mathbf{k}(\epsilon-\omega)\right]
\end{equation}

We therefore get
\begin{equation}
\kappa = \frac{N}{T}\int \frac{d^dk}{(2\pi)^d} v_\mathbf{k}^2\int d\epsilon  \frac{dn_F(\epsilon)}{d\epsilon}\epsilon^2A(\mathbf{k},0)^2
\end{equation}
Since $T\ll E_F$, we use the Sommerfeld expansion to get

\begin{equation}
\kappa_A \sim N T\int \frac{d^dk}{(2\pi)^d} v_\mathbf{k}^2A(\mathbf{k},0)^2\sim N\frac{\nu\langle v_{el}^2\rangle}{\lambda}.
\end{equation}

\subsubsection{Phonon drag}
Momentum is not even approximately conserved in our model, as umklapp scattering by a phonon into the second Brillioun zone is just as probable as a normal scattering process. Therefore, we need not worry about a sharp Drude peak due to an approximate sound mode in the system \cite{HartnollReview}. However, the occupation numbers of the $N^2$ phonons are an approximately conserved quantity, as their lifetime is finite only due to $1/N$ effects (see Eq. (\ref{eq:Pi}))
. Therefore, their effect on the thermal conductivity must be carefully studied; this is known as phonon drag \cite{Holstein}.

The diagrams which represent phonon drag in thermal transport are those that may be separated into two parts by severing two phonon propagators. The leading order diagram is given in Fig. \ref{fig:diagrams2}. These diagrams are naively suppressed by $1/N$; however, as was shown before, the phonon propagator pair $D^R(\epsilon)D^A(\epsilon+\omega)\approx \delta(\epsilon\pm\omega_0)/[\omega+i\Gamma]$ gives a contribution which is $O(N)$ in the d.c. limit, $\omega\rightarrow 0$.

\begin{figure}	
  \centering 
\includegraphics[width=0.5\textwidth]{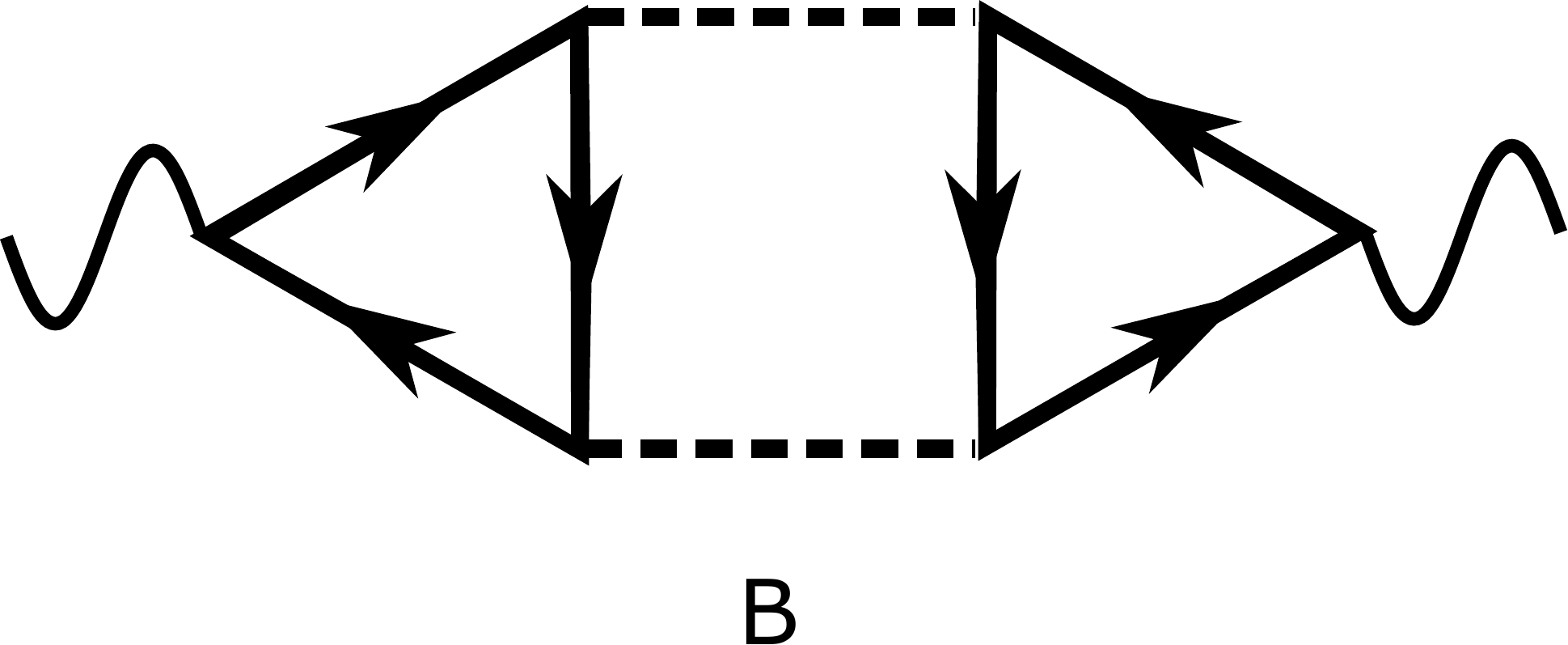}
  \caption{The leading order diagram which represents phonon drag. Arrows are fully dressed electron propagators, dashed lines are phonons, and the squiggly line represents $J_E$.}
\label{fig:diagrams2}
\end{figure}

We calculate the contribution of this diagram following \cite{Carrega}. It is given by
\begin{equation}\label{PiB}
\Pi_B(i\omega_n) = \frac{1}{\beta}\sum_{\Omega_n}\int \frac{d^dq}{(2\pi)^d}\Gamma_\mathbf{q}(i\Omega_n,i\Omega_n+i\omega_n)\Gamma_\mathbf{q}(i\Omega_n+i\omega_n,i\Omega_n)D_\mathbf{q}(i\Omega_n)D_\mathbf{q}(i\Omega_n+i\omega_n)
\end{equation}
where 
\begin{equation}
\begin{split}
\Gamma_\mathbf{q}(i\omega_1,i\omega_2) = \frac{\alpha^2}{M\omega_0\beta}\sum_{\nu_n} \nu_n\int \frac{d^dk}{(2\pi)^d} &\left[v_{\mathbf{k+q}}G_\mathbf{k}(i\nu_n)G_\mathbf{k+q}(i\nu_n+i\omega_1)G_\mathbf{k+q}(i\nu_n+i\omega_2)\right.\\
&+\left. v_{\mathbf{k-q}}G_\mathbf{k}(i\nu_n)G_\mathbf{k-q}(i\nu_n-i\omega_1)G_\mathbf{k-q}(i\nu_n-i\omega_2)\right]
\end{split}
\end{equation}

The function
\begin{equation}
f(z) = \Gamma_\mathbf{q}(z,z+i\omega_n)\Gamma_\mathbf{q}(z+i\omega_n,z)D_\mathbf{q}(z)D_\mathbf{q}(z+i\omega_n)
\end{equation}
has a branch cut when $\Im[z] = 0,-\omega_n$ and $\Re[z]\ne 0$; therefore, the Matsubara summation over $\Omega_n$ in Eq. (\ref{PiB}) results in (after analytical continutation $i\omega_n\rightarrow \omega+i\delta$)
\begin{equation}
\begin{split}
\Pi_B(i\omega_n) &= \int \frac{d^dq}{(2\pi)^d}\int \frac{d\Omega}{2\pi i}\left[n_B(\Omega)-n_B(\Omega+\omega)\right]\\
&\times\Gamma_\mathbf{q}(\Omega-i\delta,\Omega+\omega+i\delta)\Gamma_\mathbf{q}(\Omega+\omega+i\delta,\Omega-i\delta)D^A_\mathbf{q}(\Omega)D^R_\mathbf{q}(\Omega+\omega)
\end{split}
\end{equation}
and therefore
\begin{equation}
\begin{split}
\kappa_B &= -\frac{1}{T} \int \frac{d^dq}{(2\pi)^d}\int \frac{d\Omega}{2\pi i}\frac{dn_B(\Omega)}{d\Omega}\Re\left[\Gamma_\mathbf{q}(\Omega-i\delta,\Omega+i\delta)\Gamma_\mathbf{q}(\Omega+i\delta,\Omega-i\delta)D^A_\mathbf{q}(\Omega)D^R_\mathbf{q}(\Omega)\right]\\
&\approx \frac{1}{T} \int \frac{d^dq}{(2\pi)^d} \frac{T}{\omega_0^2} \Gamma_\mathbf{q}(\omega_0-i\delta,\omega_0+i\delta)\Gamma_\mathbf{q}(\omega_0+i\delta,\omega_0-i\delta)\frac{1}{\Gamma}\\
&=\frac{N}{T}\frac{T^2}{\omega_0^4}\int \frac{d^dq}{(2\pi)^d}\Gamma_\mathbf{q}(\omega_0-i\delta,\omega_0+i\delta)\Gamma_\mathbf{q}(\omega_0+i\delta,\omega_0-i\delta)
\end{split}
\end{equation}
where we have used $\frac{dn_B(\Omega)}{d\Omega} =-\frac{T}{\Omega^2}$ for $T\gg \Omega$.

$(M\omega_0/\alpha^2)\Gamma_\mathbf{q}(\omega_0-i\delta,\omega_0+i\delta)$ is given by (see \cite{Carrega} for details):

\begin{equation}
\begin{split}
\int \frac{d^dk}{(2\pi)^d} v_\mathbf{k+q}\int \frac{d\epsilon}{2\pi}&\left[\epsilon\left[n_F(\epsilon)-n_F(\epsilon+\omega_0)\right]A_\mathbf{k}(\epsilon)G^A_{\mathbf{k+q}}(\epsilon+\omega_0)G^R_{\mathbf{k+q}}(\epsilon+\omega_0)\right.\\
&-i\left.(\epsilon-\omega_0)n_F(\epsilon)G^R_\mathbf{k}(\epsilon-\omega_0)G^R_{\mathbf{k+q}}(\epsilon)G^R_{\mathbf{k+q}}(\epsilon)+c.c.\right]\\
+\int \frac{d^dk}{(2\pi)^d} v_\mathbf{k-q}\int \frac{d\epsilon}{2\pi}&\left[\epsilon\left[n_F(\epsilon)-n_F(\epsilon-\omega_0)\right]A_\mathbf{k}(\epsilon)G^A_{\mathbf{k-q}}(\epsilon-\omega_0)G^R_{\mathbf{k-q}}(\epsilon-\omega_0)\right.\\
&-i\left.(\epsilon+\omega_0)n_F(\epsilon)G^R_\mathbf{k}(\epsilon+\omega_0)G^A_{\mathbf{k-q}}(\epsilon)G^R_{\mathbf{k-q}}(\epsilon)+c.c.\right]
\end{split}
\end{equation}
in the last two lines, we change $q\rightarrow -q$, which results in an additional minus signs, as the two integrals are odd in $q$:

\begin{equation}
\begin{split}
\Gamma_\mathbf{q}(\omega_0-i\delta,\omega_0+i\delta) \\
 = \int \frac{d^dk}{(2\pi)^d} v_\mathbf{k+q}\int \frac{d\epsilon}{2\pi}&\left[\epsilon\left[n_F(\epsilon)-n_F(\epsilon+\omega_0)\right]A_\mathbf{k}(\epsilon)G^A_{\mathbf{k+q}}(\epsilon+\omega_0)G^R_{\mathbf{k+q}}(\epsilon+\omega_0)\right.\\
&+\epsilon\left[n_F(\epsilon-\omega_0)-n_F(\epsilon)\right]A_\mathbf{k}(\epsilon)G^A_{\mathbf{k+q}}(\epsilon-\omega_0)G^R_{\mathbf{k+q}}(\epsilon-\omega_0)\\
&-i(\epsilon-\omega_0)n_F(\epsilon)G^R_\mathbf{k}(\epsilon-\omega_0)G^R_{\mathbf{k+q}}(\epsilon)G^R_{\mathbf{k+q}}(\epsilon)+c.c.\\
&\left.+i(\epsilon+\omega_0)n_F(\epsilon)G^R_\mathbf{k}(\epsilon+\omega_0)G^R_{\mathbf{k+q}}(\epsilon)G^R_{\mathbf{k+q}}(\epsilon)+c.c.\right].
\end{split}
\end{equation}

To lowest order in $\omega_0/\sqrt{gT/\nu}$, only the last two lines contribute; this results in (expanding $G_\mathbf{k}(\epsilon\pm\omega_0)$ to lowest order in $\xi_\mathbf{k}/\sqrt{gT/\nu}$ so the momentum integral does not vanish)
\begin{equation}
\begin{split}
&\Gamma_\mathbf{q}(\omega_0-i\delta,\omega_0+i\delta)\\
&\approx-i\frac{\omega_0^2}{T}\frac{gT}{\nu} \int \frac{d^dk}{(2\pi)^d} v_\mathbf{k+q}\xi_\mathbf{k}\int \frac{d\epsilon}{2\pi}\epsilon n_F(\epsilon)\left(\frac{1}{\epsilon+i\sqrt{8gT/\nu-\epsilon^2}}\right)^5\left[1-i\frac{\epsilon}{\sqrt{8gT/\nu-\epsilon^2}}\right]+c.c.\\
&\sim \frac{\omega_0^2}{T} \left(\frac{\nu}{gT}\right)^{1/2}\int \frac{d^dk}{(2\pi)^d} v_\mathbf{k+q}\xi_\mathbf{k}
\end{split}
\end{equation}

This results in a phonon-drag thermal conductivity
\begin{equation}\label{eq:kappaB}
\kappa_B \sim N\frac{\langle v_{el}^2\rangle }{T}\frac{E_F}{gT}\sim N\frac{\nu\langle v_{el}^2\rangle }{\lambda}\left(\frac{E_F}{T}\right)^2\gg \kappa_A.
\end{equation}

Higher order phonon-drag diagrams, like the one depicted in Fig. \ref{fig:diagrams3} are suppressed by factors of $E_F/gT$ relative to the lowest order diagram. Therefore, $\kappa\approx\kappa_B$.

\begin{figure}	
  \centering 
\includegraphics[width=0.5\textwidth]{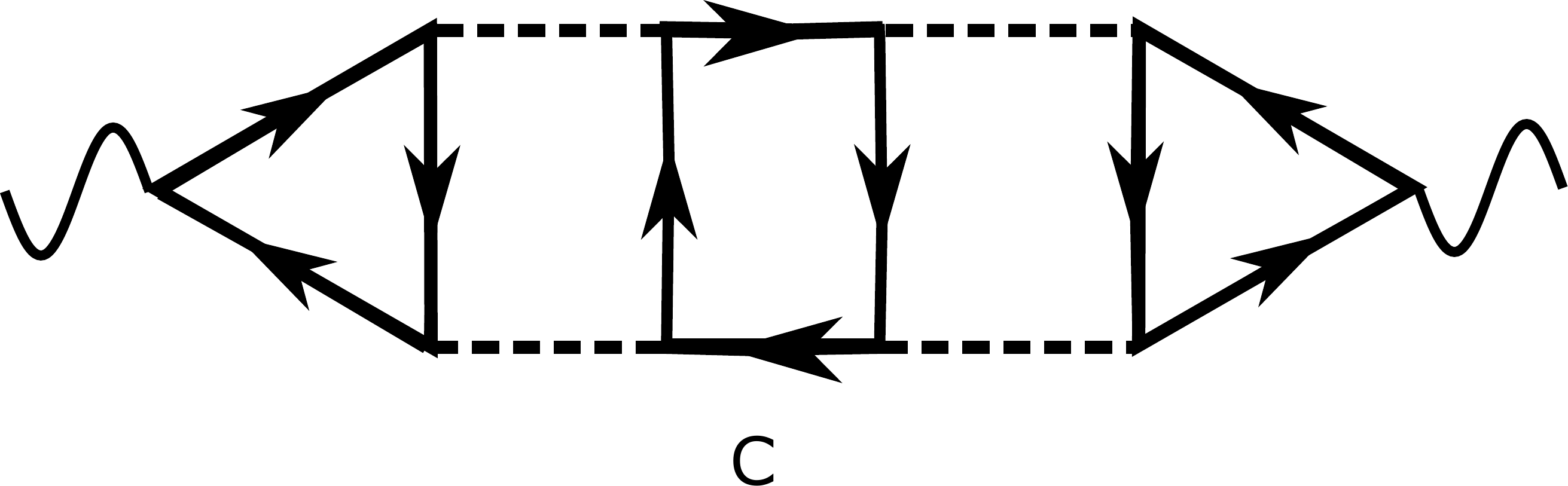}
 \caption{A subleading phonon drag diagram. Arrows are fully dressed electron propagators, dashed lines are phonons, and the squiggly line represents $J_E$. This diagram is suppressed by a factor of $E_F/gT$ relative to diagram $B$.}
\label{fig:diagrams3}
\end{figure}

\section{Charge conductivity}
The conductivity of this model has been studied in \cite{Werman, Werman2}. The current operator is given by
\begin{equation}
J = \sum_a\int \frac{d^dk}{(2\pi)^d} v_\mathbf{k}c^\dagger_a(\mathbf{k})c_a(\mathbf{k}).
\end{equation}

We calculate the conductivity using the Kubo formula,
\begin{equation}
\sigma = -\lim_{\omega\rightarrow 0}\frac{\Im\left[\Pi^{JJ}(\omega)\right]}{\omega},
\end{equation}
with $\Pi^{JJ}(\omega)$ the retarded current-current correlation function. the naive leading order in $N^{-1}$ and in $\omega_1/\omega_0$ contribution to the conductivity is given by (depicted in Fig.~\ref{fig:diagrams1}, with the appropriate current vertex; all vertex corrections vanish for displersionless phonons)
\begin{equation}
\Pi_A^{JJ}(i\omega_n) = \frac{1}{\beta}\sum_{\nu_n}\int \frac{d^dk}{(2\pi)^d} v_\mathbf{k}^2 G_\mathbf{k}(i\nu_n)G_\mathbf{k}(i\nu_n+i\omega_n)
\end{equation}
and therefore
\begin{equation}\label{eq:sigmaB}
\sigma_A = \int \frac{d^dk}{(2\pi)^d} v_\mathbf{k}^2 A^2(\mathbf{k},0) = \frac{\nu\langle v_{el}^2\rangle}{gT}
\end{equation}

The phonon drag contribution is shown schematically in Fig. \ref{fig:diagrams2}; the leading order (in $E_F/g T$) is given by
\begin{equation}
\begin{split}
\sigma_B &=  -\int \frac{d^dq}{(2\pi)^d}\int \frac{d\Omega}{2\pi i}\frac{dn_B(\Omega)}{d\Omega}\Re\left[\Lambda_\mathbf{q}(\Omega-i\delta,\Omega+i\delta)\Lambda_\mathbf{q}(\Omega+i\delta,\Omega-i\delta)D^A_\mathbf{q}(\Omega)D^R_\mathbf{q}(\Omega)\right]\\
&\approx \int \frac{d^dq}{(2\pi)^d} \frac{T}{\omega_0^2} \Lambda_\mathbf{q}(\omega_0-i\delta,\omega_0+i\delta)\Lambda_\mathbf{q}(\omega_0+i\delta,\omega_0-i\delta)\frac{1}{\Gamma}\\
&=N\frac{T^2}{\omega_0^4}\int \frac{d^dq}{(2\pi)^d}\Lambda_\mathbf{q}(\omega_0-i\delta,\omega_0+i\delta)\Lambda_\mathbf{q}(\omega_0+i\delta,\omega_0-i\delta)
\end{split}
\end{equation}
where 
\begin{equation}
\begin{split}
\Lambda_\mathbf{q}(\omega_0-i\delta,\omega_0+i\delta) \\
 = \int \frac{d^dk}{(2\pi)^d} v_\mathbf{k+q}\int \frac{d\epsilon}{2\pi}&\left[\left[n_F(\epsilon)-n_F(\epsilon+\omega_0)\right]A_\mathbf{k}(\epsilon)G^A_{\mathbf{k+q}}(\epsilon+\omega_0)G^R_{\mathbf{k+q}}(\epsilon+\omega_0)\right.\\
&+\left[n_F(\epsilon-\omega_0)-n_F(\epsilon)\right]A_\mathbf{k}(\epsilon)G^A_{\mathbf{k+q}}(\epsilon-\omega_0)G^R_{\mathbf{k+q}}(\epsilon-\omega_0)\\
&-in_F(\epsilon)G^R_\mathbf{k}(\epsilon-\omega_0)G^R_{\mathbf{k+q}}(\epsilon)G^R_{\mathbf{k+q}}(\epsilon)+c.c.\\
&\left.+in_F(\epsilon)G^R_\mathbf{k}(\epsilon+\omega_0)G^R_{\mathbf{k+q}}(\epsilon)G^R_{\mathbf{k+q}}(\epsilon)+c.c.\right].
\end{split}
\end{equation}

The fact that the electron Green's function is momentum independent to leading order in $E_F/g T$ renders the phonon drag contribution 
\begin{equation}
\sigma_B = \frac{E_F}{g T}\sigma_A,
\end{equation}
and we neglect it. Hence, we get that $\sigma\approx\sigma_A$. Comparing Eqs. (\ref{eq:sigmaB},\ref{eq:kappaB}) we see that in the high temperature regime of our model, the Wiedemann-Franz law
\begin{equation}
\frac{\kappa}{\sigma} = \frac{\pi^2}{3}T
\end{equation}
is strongly violated.

\end{widetext}

\bibliography{Bibliography}
\end{document}